\documentclass[12pt,preprint]{aastex}
\usepackage{emulateapj5}

\tighten
\input psfig.tex

\slugcomment{Accepted for publication in ApJ}
\newcommand{\mpc}{h_{75}^{-1}\ {\rm Mpc}}

\newcommand{\ltsima}{$\; \buildrel < \over \sim \;$}
\newcommand{\lsim}{\lower.5ex\hbox{\ltsima}}
\newcommand{\gtsima}{$\; \buildrel > \over \sim \;$}
\newcommand{\gsim}{\lower.5ex\hbox{\gtsima}}
\newcommand{\eg}{{\it e.g.,\ }}
\newcommand{\ie}{{\it i.e.,\ }}

\slugcomment{Accepted for publication in {\it The Astrophysical Journal}}
\shortauthors{Postman et al.}
\shorttitle{The KPNO/Deeprange Distant Cluster Survey}

\received{2002 May 14}
\begin{document}

\title{The KPNO/Deeprange Distant Cluster Survey: I. The Catalog \&
the Space Density of Intermediate Redshift Clusters }

\author{Marc Postman\altaffilmark{1}}
\affil{Space Telescope Science Institute\altaffilmark{2},
Baltimore, MD 21218}

\author{Tod R. Lauer}
\affil{National Optical Astronomy Observatories\altaffilmark{3},
Tucson, AZ 85726}

\author{William Oegerle\altaffilmark{1}}
\affil{Laboratory for Astronomy and Solar Physics, Code 681, 
NASA/GSFC, Greenbelt, MD 20771} 

\author{Megan Donahue\altaffilmark{1}}
\affil{Space Telescope Science Institute\altaffilmark{2},
Baltimore, MD 21218}

\vfill

\altaffiltext{1}{Visiting Astronomer Kitt Peak National Observatory,
NOAO.}
\altaffiltext{2}{The Space Telescope Science Institute
is operated by the Association of Universities for
Research in Astronomy (AURA), Inc., under National Aeronautics and
Space Administration (NASA) Contract NAS 5-26555.}
\altaffiltext{3}{The National Optical Astronomy Observatories are
operated by AURA, Inc., under cooperative agreement with the National
Science Foundation.}       

\begin{abstract}
We have conducted an automated search for galaxy clusters within
a contiguous 16 square degree $I-$band survey in the north Galactic
hemisphere. A matched filter detection algorithm 
identifies 444 cluster candidates in the
range $0.2 \lsim z \lsim 1.2$. The full catalog is presented along
with the results from a follow-up spectroscopic survey. 
The estimated redshift distribution of the cluster candidates 
is consistent with a constant comoving density over the range
$0.2 \le z_{est} \le 0.8$. A decline in the cluster space density by more than
a factor of 3 over this redshift range is rejected at $>99.9$\% confidence level. 
We find that the space density of $\Lambda_{CL} \ge 40$ clusters in our survey lies 
in the range $(1.6 - 1.8) \times 10^{-5}h_{75}^{3}$ Mpc$^{-3}$, $\sim1.5$
times higher than the local distribution of comparably rich 
Abell RC $\ge 0$ clusters.  The $\Lambda_{CL}$ distribution 
is consistent with a power-law.
The discrepancy between the space density of Abell clusters and 
the clusters in this survey declines quickly as $\Lambda_{CL}$ increases,
suggesting that the difference at lower richness is due to 
significant incompleteness in the Abell catalog.
A percolation analysis reveals that 10 - 20\% of the spectroscopically
confirmed distant clusters are linked into superclusters at overdensities
between $10 < {\delta\rho / \rho} < 50$, similar to what is seen in the local 
cluster distribution. 
This suggests that there has been little evolution of the cluster-cluster 
correlation length for $z \lsim 0.5$.
\end{abstract}


\section{Introduction} 

The evolution of clustering is an inevitable consequence of any gravitationally
driven model of structure formation. The scale dependence and amplitude of this
evolution place significant constraints on the nature of
dark matter and key cosmological parameters. Clusters of galaxies provide
an efficient way to trace matter on large-scales and significant
work has been done to both measure the current epoch cluster-cluster correlation
length (\eg \citealt{bs83, p92, n92, cr97, b99, cc2000})
and to determine the constraints these measurements
place on structure formation models 
(\eg \citealt{eke, mo, col, mos01}).
In light of the extensive observational
constraints available on the clustering properties of clusters at $z < 0.1$, it
is highly desirable to obtain similar measurements at higher redshifts ($z \lsim 1$)
to constrain the evolution of clustering on large-scales and at high
masses. Indeed, the evolution of clustering of the high-mass end of the 
matter distribution is typically the most sensitive to variations in cosmological
parameters \citep{peeb}. To achieve accurate constraints on the evolution
of clustering on large-scales requires galaxy and cluster surveys that are
both deep and wide. The depth must be sufficient to sample sources significantly 
fainter than the characteristic luminosity at redshifts where significant evolution 
can be detected. The area must be large enough
to reliably sample structures on scales 
much greater than the expected correlation lengths. 

The advent of large format CCD detectors, CCD mosaic cameras, and multi-object
spectrographs has made mapping the large-scale distribution of clusters 
and galaxies at intermediate redshifts quite feasible. 
We completed the first moderately deep imaging survey that also subtended 
a substantial contiguous area ($\sim 16$ deg$^2$) to establish a database capable of 
accurately measuring the clustering of galaxies and clusters up to scales of $\sim130\mpc$
over the redshift range $0.2 \lsim z \lsim 1$, using the 
first of the large format detectors at the Kitt Peak National Observatory
\citep{dr1, dr2}.
There are now at least 5 additional optical/NIR surveys with total areas in excess of 
10 square degrees that have depths sufficient to detect a substantial number
of clusters up to redshifts near unity: 
the ESO Imaging Survey ($\sim15\ {\rm deg}^2$; \citealt{eis}), 
the NOAO Deep Wide-Field Survey ($\sim18\ {\rm deg}^2$; \citealt{ndwfs}), 
the BTC Survey ($\sim40\ {\rm deg}^2$; \citealt{btc40}), 
the Red Sequence Cluster Survey ($\sim100\ {\rm deg}^2$; \citealt{gy00}), and 
the Las Campanas Distant Cluster Survey ($\sim130\ {\rm deg}^2$; \citealt{gnz}).
X-ray cluster surveys (\eg \citealt{h92, sch97, ros98, bcsx, rom00})
have also discovered a moderate number of 
intermediate redshift ($0.5 < z < 1$) cluster over areas in excess of 100 square
degrees.  Combined with the application of a variety of objective cluster detection
algorithms, which enable the inevitable selection biases to be properly
quantified, these surveys are now yielding important and accurate constraints on
cosmological parameters (\eg $\sigma_8, \Omega_m$) as well as on the abundance,
clustering, and properties of galaxy clusters
over the last half of the current age of the universe. 

In this paper, we present the results
of our automated cluster search using the Deeprange survey along 
with measurements of the space density of clusters as a function
of redshift and richness. The scientific requirements that
were used to design the survey and a summary of the imaging observations
and data reduction are presented in \S\ref{survey}. 
The construction and calibration of the galaxy catalog used as 
the input for the cluster detection algorithm
are described in \S\ref{galcat}.
The algorithm and resulting cluster catalog are presented in \S\ref{cluscat}
along with a detailed description of the derivation of the selection
function, richness calibration, and false positive rate estimates.
We have also performed a follow-up spectroscopic survey of the
cluster candidates, which is described along with the spectroscopic
data reduction methods in \S\ref{specfu}. The main results of this
paper are presented in \S\ref{results} and are followed by a summary
of our key conclusions in \S\ref{conc}. Throughout this paper we define
$h_{75} = {\rm H}_{\circ}/(75$ km s$^{-1}$ Mpc$^{-1})$ and adopt
$\Omega_m = 0.2$, $\Omega_{\Lambda} = 0$.

\section{Survey Design}
\label{survey}

Our survey field is a contiguous $4^{\circ} \times 4^{\circ}$ area located
at high Galactic latitude. This geometry is motivated by a number of important
considerations.
First, to explore large-scale clustering of clusters at $z \sim 1$, the
survey must subtend a contiguous comoving scale that is larger than
50$\mpc$ -- the locally observed zero-crossing scale in the cluster-cluster two-point 
correlation function \citep{kr}. Indeed, the linear comoving scale of the 
survey must at least be comparable with $130\mpc$ -- 
the size of the largest structures in the local galaxy distribution 
(\eg \citealt{v92, lcrs, col2001, z2002}).
At $z = 1$, 130$\mpc$ (comoving) corresponds to an
angular scale of $\sim2.7^{\circ}$. Second, to measure low-order
clustering statistics at intermediate redshifts with accuracies
comparable to those at low redshift our survey needs to subtend a volume 
that is comparable with the typical low redshift cluster survey. 
A $\vert b\vert \ge 30^{\circ}$ cluster survey to $z = 0.07$ (\eg \citealt{bs83})
subtends approximately $4 \times 10^7 h_{75}^{-3}$ Mpc$^3$ -- 
to cover a similar volume out to $z = 1$ would require a solid angle
of approximately 20 deg$^2$. We also need to have approximately 300 to 500
clusters in the survey area ($\sim$25\%\ of which would 
lie within $0.75 \leq z \leq 1.15$)
to assure moderately accurate measurements of the clustering at higher redshifts.
When we began to plan this survey, estimates of the surface density 
of clusters from small area surveys \citep{gunn, p96}
were in the range 15-20 clusters deg$^{-2}$ out to $z = 1,$ and thus 
a survey area of at least 16 deg$^2$ was required.
Lastly, although one can construct a survey that samples these scales by combining
data from many non-contiguous fields, finite sampling and cosmic
variance effects can severely degrade the accuracy of clustering statistics derived
from such a patchwork survey (cf. \citealt{dr1}). 

\vspace{\baselineskip}
\epsfxsize=\hsize
\centerline{\epsfbox{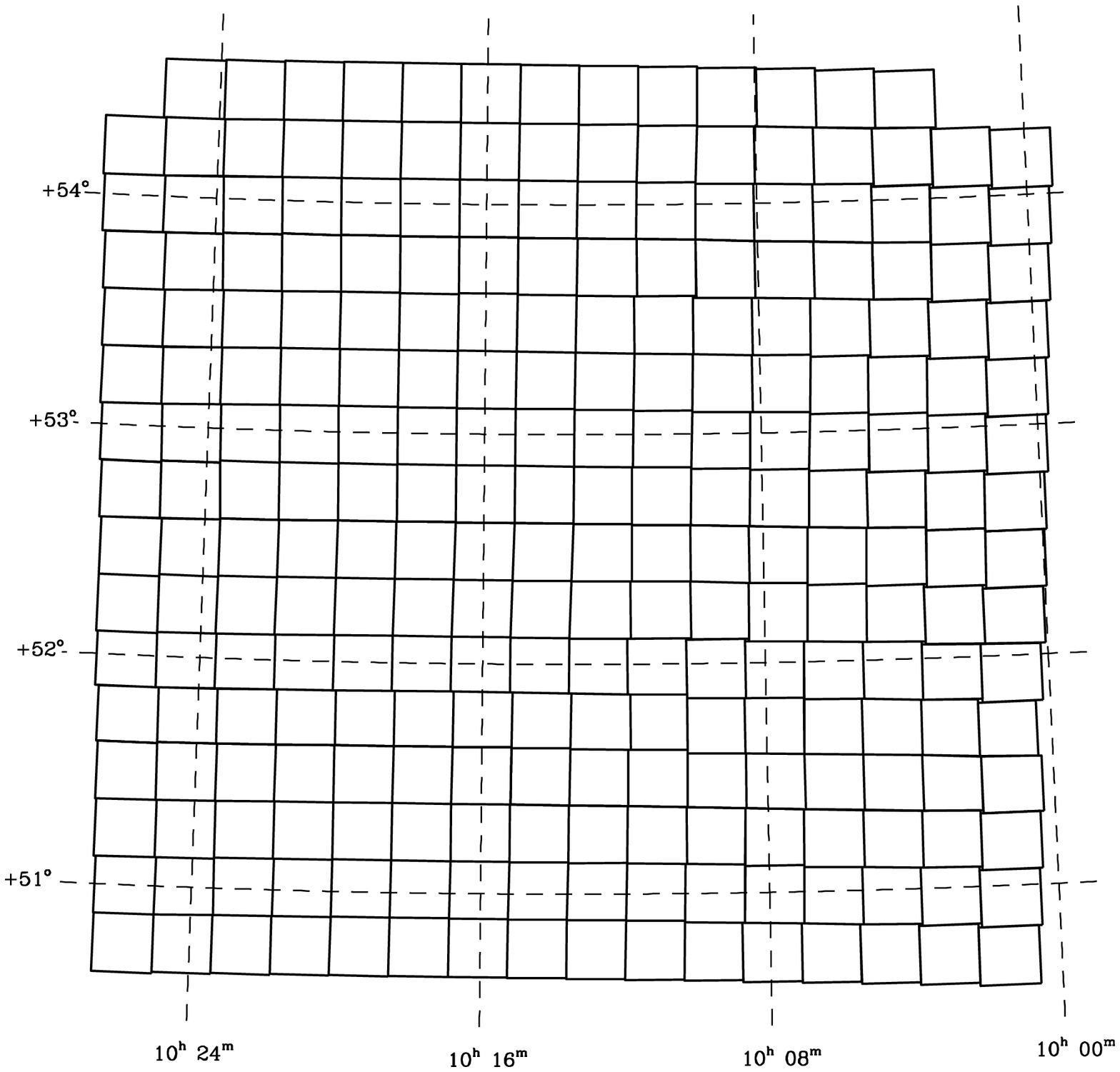}}
\figcaption{\footnotesize
The positions of the 253 survey pointings on the sky.
Grid 001 is located at the southeastern corner. The survey originally
was designed with 256 pointings but 1 pointing was excluded due
to the presence of a V = 7 mag star (HD 90249) and 2 others were never observed
due to weather limitations. Each 16 arcminute grid overlaps its adjacent grids by
1 arcminute enabling accurate photometric and astrometric calibration
across the entire survey.
\label{figsmap}}
\vspace{\baselineskip}

The survey is centered at 10h 13m 27.95s +52d 36m 43.5s (J2000)
by virtue of its high Galactic latitude ($+51^{\circ}$),
low HI column density ($2.2 \times 10^{20}\ {\rm cm}^{-2}$),
high declination (enabling extended visibility from KPNO), low
IRAS $100\mu$ cirrus emission, and the absence of many bright stars or 
nearby rich clusters. The E(B-V) estimates from \citet{sch98} (computed
using the NED extinction calculator) for our survey
region yield a mean value of 0.011 mag, a maximum value of 0.018 mag, and a minimum value
of 0.005 mag. The mean E(B-V) value of 0.011 mag corresponds to
an $I-$band extinction of 0.02 mag (assuming an $R_V=3.1$ extinction curve).
 
\subsection{Observations}

The imaging phase of the survey was conducted in the $I-$band using the
prime focus CCD camera on the KPNO Mayall 4m telescope. Working in the $I-$band assures
good completeness in the detection of clusters out to 
$z \sim 0.8$ (see \S\ref{selfunc}). The camera has a $16'$ field
of view (0.47 arcseconds pixel$^{-1}$ for the T2KB CCD); 
256 exposures were, therefore, required
to survey the entire field. Each pointing overlapped its adjacent pointing by 
1 arcminute -- the overlap is required to guarantee accurate photometric
and astrometric calibration of the galaxy catalog (see \S\ref{pcal} and \S\ref{acal}).
Figure~\ref{figsmap} shows the configuration of the survey pointings on the sky.

The observations were conducted over a 2 year period between Jan. 1994 and March 1996. 
Each exposure was 900 seconds in duration\footnote{CR splits were not performed because
the cosmic ray hit rate is low -- no significant increase in spurious
detections or contamination of object photometry was found. The overheads associated with
CR splits are substantial.}. This allows us to
reach a $5\sigma$ detection limit (for point sources) of I=23.85, sufficient to
detect cluster galaxies 2 magnitudes fainter than the typical first-ranked elliptical 
at $z = 1$ (\eg \citealt{abk, pol}).
Going to this depth was essential -- a shallower survey would only be
sufficient for detecting the very richest $z \sim 1$ clusters and would limit our
conclusions about evolution of structure. (At I = 23.5, we are able to
detect $z \sim 1$ Abell richness class 1 systems). Since we also wish to
conduct follow-up spectroscopic surveys based on these images, it was
desirable to have the data reach the spectroscopic limit of 8 - 10m class
telescopes with a reasonable signal-to-noise ratio.
The median seeing achieved in the images is 1.3 arcseconds FWHM (85\% of the
images have seeing of 1.5$''$ or better) and the median airmass is 1.2. 
Although the survey was designed with 256 pointings, only 253 images were ultimately
acquired -- 1 pointing was excluded due to the presence of a V = 7 mag star (HD 90249) 
and 2 others were never observed due to weather limitations.

Fringing was minimized by using the prime focus camera scan table.
The scan table suppresses flat-field and fringing artifacts by 
physically moving the detector in sync with and parallel to
an electronic shift of charge.
After some experimentation, it was determined that a 60-pixel scan was the
optimal choice (the fringe amplitude decays more slowly with scan
size when the scan size exceeds 60 pixels but decays quite rapidly
with scan size for scans less than 60 pixels). With this set up, we were able
to flatten the images to 1 percent or better. The scan process reduces the usable
field of view in the scan direction by 28 arcseconds. The reduced images from this
survey are publicly available from the 
NOAO science archive (http://archive.noao.edu/nsa/).

\subsection{Imaging Data Reduction}

To prepare the image data for use in galaxy and cluster detection,
all significant instrumental signatures are removed using well-tested
data reduction procedures. Flat-fielding and de-fringing of the image 
data represent the most significant steps in this process. 
We begin the reduction process with
basic de-biasing and removal of pixel-to-pixel sensitivity
variations using an averaged dome flat image.
We then create a master median sky image using
all the images from a given observing run. An illumination correction image
(intended to remove large-scale sensitivity variations)
is derived from this median image by fitting a two-dimensional, second-order surface
and then normalizing the mean pixel value in the fit to unity. 
A fringe correction image is derived by dividing the master median
by the illumination correction image.
The illumination and fringe correction processes are then performed
in the usual manner.
The resultant images exhibit extremely uniform sky levels: typical
sky variations are less than 0.2\% on scales between 50$''$ and 240$''$.
Furthermore, small (1 pixel) to medium (30 pixels) scale sky variations 
are consistent with photon statistics indicating that the fringe 
removal has worked extremely well.

\section{The Galaxy Catalog}
\label{galcat}

We use a modified version of the
Faint Object Classification and Analysis System (FOCAS; \citealt{jt81, v82})
to detect and classify objects in the images. The 
modifications include an algorithm that allows the use of a variable PSF in an image 
for object classification and an algorithm that cleans spurious detections
surrounding bright stars and galaxies.
We use a $3-\sigma$ detection threshold\footnote{In calibrated units, the
detection threshold is typically $\mu_{I,det} = 24.2$ mag arcsec$^{-2}$ but,
of course, this depends on the mean sky level.} and a 12 pixel (2.65 arcsec$^2$) minimum 
object area constraint. These parameters work well for 95\% 
complete detection of galaxies with isophotal magnitudes that are 
0.5 mag brighter than the detection threshold 
(\eg a 95\% completeness limit of $I=23.5$ when $\mu_{I,det} = 24$). 

We run FOCAS on each individual $16' \times 16'$ image and construct
a corresponding instrumental object catalog. These individual catalogs
are then merged to construct our master galaxy catalog once the photometric 
and astrometric calibration of the entire survey is completed.
We manually defined $\sim4000$ rectangular regions in the vicinity of bright
stars or galaxies ($I \lsim 14$) where object detection is compromised
due to excessive pixel saturation, high intensity PSF wings, and/or 
significant scattered light. Any detections that lie within these regions
are excluded from the analyses. The total survey area is 14.7 deg$^2$ after
accounting for all the exclusion regions.

\subsection{Star -- Galaxy Classification}
\label{sgclass}

The FOCAS object classifier attempts a 2D fit of
a linear combination of a stellar PSF and a broadened PSF\footnote{
The broadened PSF is simply a wider version of the stellar PSF. See
equation 10 in \citet{p96} for details.} 
for each detected object. The classification
of an object depends on what fraction of the fit is represented by
the broadened PSF. Typically, if a PSF broadened by more than
a factor of 1.3 comprises at least 25\% of the amplitude of 
final fit the object is classified as a galaxy. The stellar PSF 
is derived by selecting $\sim25$ star-like sources in each image.

The images in this survey were obtained prior to the installation
of the enhanced prime focus corrector (installed in late 1996). 
Hence, there are noticeable PSF
variations from the image center to the edge. The standard FOCAS
resolution classifier uses a position invariant PSF. We modified
this software to allow a position dependent PSF model to be employed.
This modification was critical in obtaining uniform object classification
across the field of view. We derive an empirical position dependent
PSF model based on measurements of the FWHM as a function of distance
from the optical center. We find that the expression
\begin{equation}
B(r) = a_0 + a_2 r^2 + a_4 r^4
\end{equation}
is an excellent representation. In this equation, $B(r)$ is the
FOCAS broadening factor and $r$ is the radial distance from the
optical center. We fit equation 1 to the lower envelope of
the FOCAS broadening factor as a function of $r$. The maximum
broadening factor allowed for an object classified as a star
is then taken to be min[$B(r) + 0.3$, 2]. The upper limit of 2 is
set by an assessment of the maximum broadening induced
\vspace{\baselineskip}
\epsfxsize=\hsize
\centerline{\epsfbox{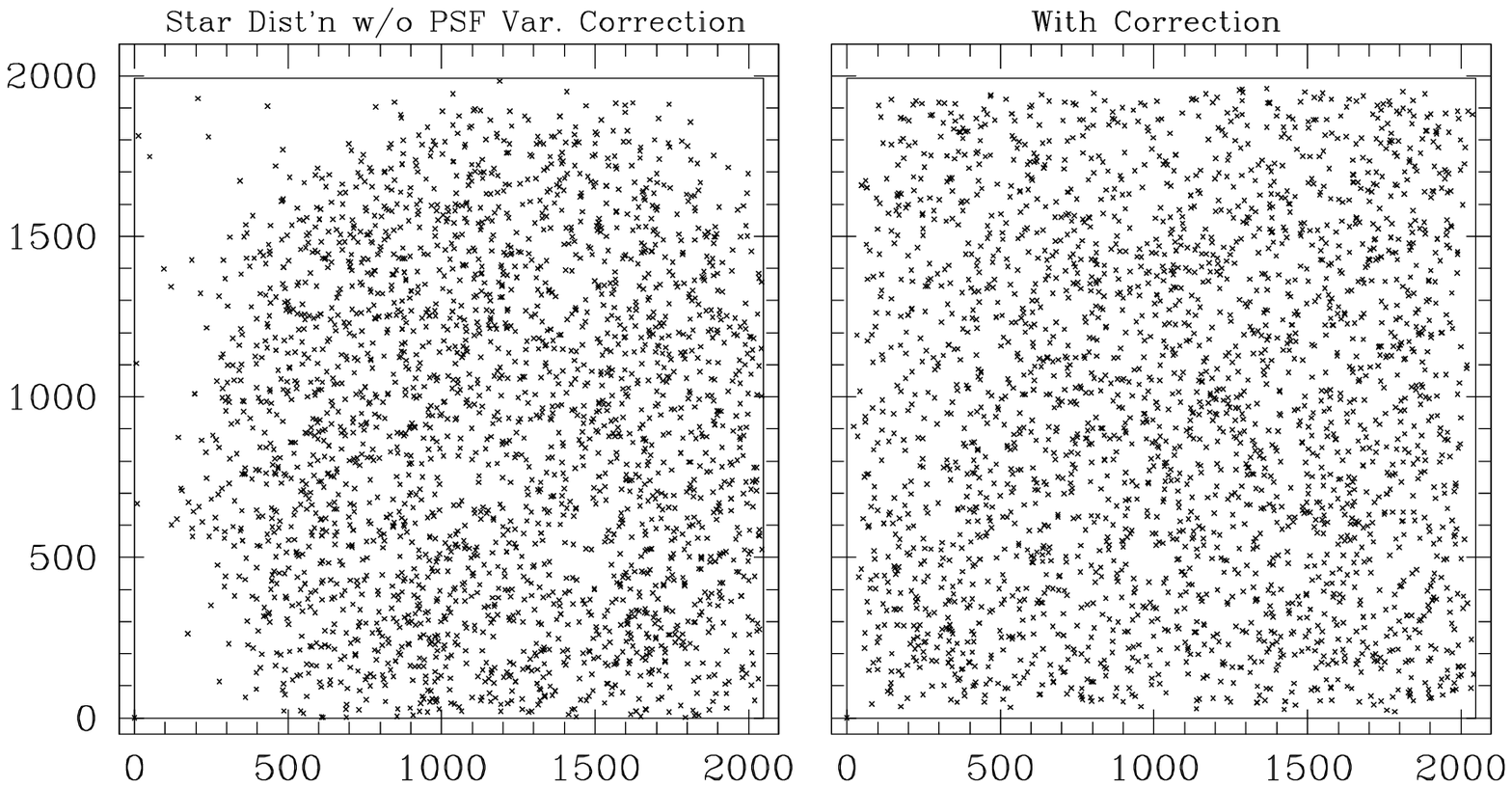}}
\figcaption{\footnotesize
The distribution of objects classified as stars before
(left) and after (right) the application of the position-dependent
PSF classifier. The data are based on a stacking of objects in 69
of the 253 images. The plot shows a random sampling of 5000 of the
$\sim60,000$ stellar objects with $19 \le I < 23$. The optical center
of the imager is located near x,y position (1200,920).
\label{figstarstack}}
\vspace{\baselineskip}
\noindent by the PSF distortion near the detector edges. The 0.3 offset is
the default used in FOCAS (\eg in an image with a constant PSF,
the default FOCAS classification rule defines stars as
objects with $B = 1.0 \pm 0.3$).
The location of best focus is not always at $r = 0$ and it can
vary from run to run. Therefore, the
coefficients are derived independently for each image and 
the $a_2$ coefficient can be positive or negative (the $a_0$ and
$a_4$ coefficients are always positive). 
In fact, the $a_2$ coefficient is negative for 98\% of the images
because the region of good focus is fairly wide and significant PSF
degradation occurs only near the edges of the CCD field of view.  
The optical center is derived by plotting the positions
of all objects classified as stars {\it prior} to the use of the
position dependent PSF. There is a well defined circular boundary
beyond which the stellar density falls off significantly (and the galaxy
surface density increases correspondingly). The center
of this region is the optical center and is located at pixel
position (1200, 920) in our data. Figure~\ref{figstarstack} shows
a random sampling of the distribution on the CCD of $\sim60,000$ objects classified 
as stars (with $19 \le I < 23$)
before and after application of the position dependent 
PSF classifier. Only minor inhomogeneities in the stellar distribution
remain after accounting for the PSF variations.
There is also no signature of the CCD image size in the galaxy two-point
correlation function \citep{dr1} providing
further validation that the above procedure for correcting the PSF variations
is successful. 

The FOCAS classifier works well down to about $I=21.5$ in these data (based
on running the same code on simulations). Fainter than this, a statistical
approach is needed to assess the probability that an object is a galaxy.
Specifically, we find that for $I > 21.5$ there is a systematic tendency for 
the classifier to identify galaxies as stars. 
We have quantified this effect by extrapolating
the bright star counts ($I \le 20$) to fainter limits and then
compute the number of stars that should be in each mag bin. From this we
generate the probability (as a function of magnitude) 
that a given faint star is really a galaxy.  That function is:
\begin{equation}
P(s \rightarrow g) = 0.663 - 5.33\times10^{-3} I^2 + 1.00\times10^{-5} I^4
\end{equation} 
where $I$ is the isophotal mag and $P(s \rightarrow g)$ is the probability that a star
of magnitude $I$ is really a galaxy. We clamp this function to lie in the range [0,1]
as it is a polynomial best fit and does not give valid
results outside the range of the fit ($18.25 \le I \le 24.5$).
The classifier rarely misclassifies
a faint star as a galaxy, however. We note that
for $I > 21.5$ galaxies outnumber stars by at least a factor of 5
(the galaxy/star ratio is $\sim10$ by $I=23.5$, based on the statistically
corrected star and galaxy counts).
Hence, star contamination would not exceed
$\sim 10 - 15$\% even if one simply classified every object fainter than
$I = 21.5$ as a galaxy. We, however, do use the above probability function
to assemble an object catalog with statistically reliable galaxy counts
down to $I=24$. Our galaxy counts as a function of $I$ magnitude are published
in \citet{dr1}. In particular, we refer the reader to figure 2 of that paper,
which shows a comparison of our galaxy counts to those from other surveys.
The good agreement between our counts and those from other surveys demonstrates
the accuracy of our statistical corrections to our faint object classifications.

\subsection{Photometric Calibration of the Survey}
\label{pcal}

Accurate reduction of the survey images to a common photometric zeropoint
is essential to ensure both uniform detection of galaxy clusters over
the area of the survey, as well as for characterizing the galaxy-galaxy
correlation function on large angular scales.
The basic calibration strategy is to use the overlap of any survey image
with its north/south and east/west neighbors in the $16\times16$ grid
of individual CCD fields to enforce local consistency of the
photometric scale.  This method is actually essential, as the entire
survey was observed in a mix of non-photometric/photometric sky
conditions, thus some fields can only be calibrated by
reference to their neighbors.  Global uniformity over the survey
is provided by sequences of fields observed in photometric conditions,
as well as a set of independent calibration observations, kindly provided
by R. Y. Shuping, using a large-field CCD camera at the KPNO 0.9m telescope.
The zeropoint, itself, is established by \citet{landolt} standards
observed at both the KPNO 4m and 0.9m telescopes.

Measuring the photometric scale of any image with respect to its
neighbors begins with identifying common objects in the catalogues of
the overlap regions.  This is done for each pair of images
that have significant angular overlap.  The second step is to
estimate the photometric offset between the two images based on the
photometric offsets between all objects in common, under an assumed
initial estimate of the photometric zeropoint.
In practice, to treat both images equally, the offset is measured as the
constant that best minimizes the differences of the {\it ad hoc} magnitudes for
each object, weighted by its average luminosity within the two images.
One can visualize this graphically as trying to fit a flat-line to a simple
plot of magnitude versus magnitude for the common objects rotated $45^\circ,$ to
avoid treating magnitudes in the two images as a pair of independent and
dependent variables.
The statistical accuracy of the photometric offset for any image
pair is determined by the density of objects in the overlap region,
and their distribution in luminosity; its median value over all pairs
in the survey is 0.019 mag.

The set of photometric differences between all neighboring images
in the survey grid defines a set of equations that can be solved
to estimate the photometric correction for any individual image
to make it consistent with the survey overall; \citet{kor} present
an independent development of this approach.
There are 253 usable images in the survey, thus
we require 253 photometric corrections, $\Delta_i.$
The photometric offsets, $m_{ij},$ between adjacent images are defined in
the sense that the magnitudes of objects in grid $j$ are subtracted
from grid $i.$  If we solve for $\Delta_i$ by a standard least-squares
procedure, then we must minimize the overall error:
\begin{equation}
E=\sum_{i=1}^{253}\sum_{j=1}^{253}\epsilon_{ij}\left(\Delta_i-m_{ij}-
\Delta_j\right)^2,
\end{equation}
where $\epsilon_{ij}=1$ when grids $i$ and $j$ are adjacent, and is
zero otherwise.  If we minimize $E$ with respect to $\Delta_i,$ then
we get:
\begin{equation}
{\partial E\over\partial\Delta_i}=0=\Delta_i\sum_{j=1}^{253}\epsilon_{ij}-
\sum_{j=1}^{253}\epsilon_{ij}m_{ij}-\sum_{j=1}^{253}\epsilon_{ij}\Delta_j.
\end{equation}
This generates a system of linear equations that can be solved
for $\Delta_i.$  The problem is over-determined as each survey row has a maximum of
15 photometric differences for $16\times15=240$ east-west comparisons
and likewise 15 sets of 16 north-south inter-row comparisons, for a maximum total
of 480 $m_{ij}$ (in fact, because there were only 253 frames available, there
are only 474 comparisons).
Monte Carlo simulations show that the error in the
photometric offsets are 0.017 mag, slightly smaller than the error
in the individual pair-wise differences.

One serious problem is that the photometric differences
between image pairs do not constrain the overall average zeropoint of the
survey.  The solution is simply to require one of the images to
have a reference zeropoint based on standard-star calibrations, the
final offset corrections are then defined to this zeropoint.
A more serious problem is that calibration based on differences leads
to the best local agreement between any image and its neighbors
in the survey, but is prone to substantial drift of the zeropoint
over large angular scales.  If the individual images
have a slight zeropoint drift over their extent, as might result
from small errors in the flat-field illumination pattern, then
this error may actually be compounded by this calibration process.
For example, if there were a 2\%\ north to south gradient in the photometric
scale of any image, then the total drift across the declination
extent of the survey would exceed 30\%\ (if all images had the same error).
The solution is to use external calibration frames plus those survey
images taken in photometric conditions with good standard star solutions
to guard against any photometric drift.  In practice, the initial
zeropoint determined from differences alone appeared to drift by 0.22 mag
in declination and 0.08 in RA over the extent of the survey, which
was readily corrected with simple linear terms.  The larger drift
in declination probably reflects the fact that
the survey was extended in declination over a number of runs, while
we were able to scan a few complete east-west rows in a single night.
We note that on photometric nights the pattern of measured
photometric differences clearly reflected the airmass variation as the
survey field transited the sky.

\subsection{Astrometric Calibration}
\label{acal}

We calibrate the astrometry using objects detected in the Digitized Palomar
Observatory Sky Survey (DSS). 
For each CCD frame, we typically cross identify 32 DSS objects (minimum
used was 10, maximum used was 60). The objects matched are chosen to
exclude saturated stars ($I \lsim 17$) and objects fainter than $I \sim 19$.
A six-term, second order astrometric solution is then derived using the CCD x-y
positions and the DSS celestial coordinates. The mean rms of the fits is
$0.56''$ and is dominated by the uncertainty of the centroids
of faint objects in the DSS images.
\vspace{\baselineskip}
\vspace{\baselineskip}
\epsfxsize=\hsize
\centerline{\epsfbox{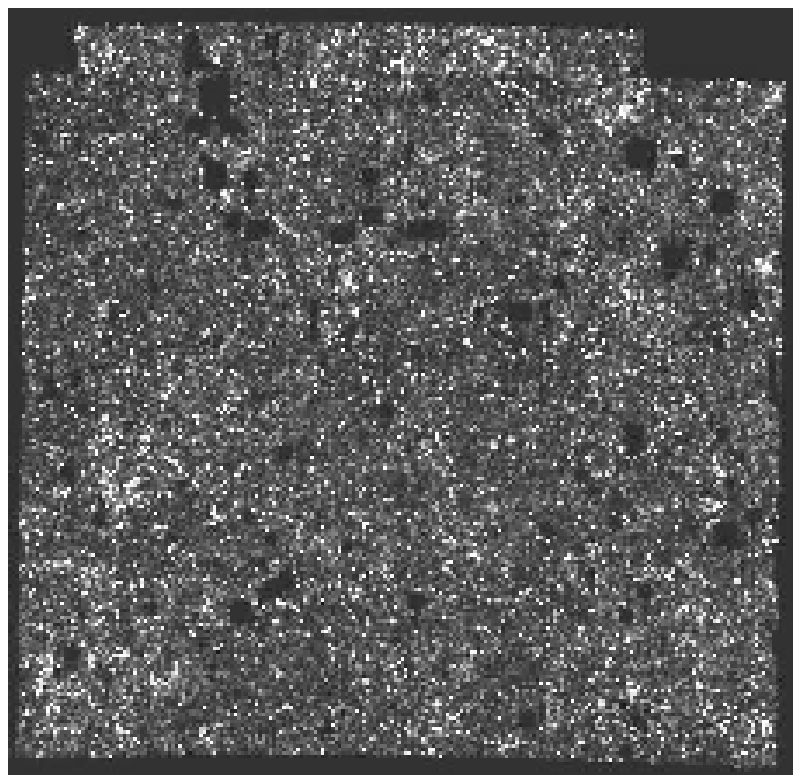}}
\figcaption{\footnotesize
The angular distribution of the $\sim710,000$ galaxies in our survey
with $I \le 23.5$. Each galaxy is represented
as a point with intensity proportional to dex(-0.4m) where m is the isophotal
$I-$band magnitude. Blank rectangular areas are the regions
excluded due to the presence of bright stars.
\label{figgalmap}}
\vspace{\baselineskip}
\noindent Systematic errors associated with the DSS astrometric calibration itself
can be as large as $2''$ but relative astrometry at the $0.3''$ level can
be achieved, as evidenced by the success of our spectroscopic
follow-up (see \S\ref{specfu}). We also derive the T2KB CCD pixel scale as
part of the astrometric calibration process and find a mean value
of $0.4702 \pm 0.0005$ arcsec pixel$^{-1}$, in
excellent agreement with the KPNO published value.

As a final check on our astrometric calibration, we cross-correlate
our catalog with the FIRST VLA survey catalog \citep{bwh95}.
We look for any image-wide
astrometric offsets as well as any systematic offsets that may be a function
of CCD position. We detect image-wide systematic offsets between our
initial DSS astrometry and that from the FIRST catalog that are typically
$0.5 - 1''$ in amplitude. These are corrected in our final catalog. Any
CCD position dependent residuals are typically $0.3''$ or less.
Once astrometric calibration is complete, we assemble a master catalog.
In cases where an object falls in an overlap region between adjacent
images, the object parameters used in the master catalog are chosen
according to where the object lies with respect to the line bisecting the 
overlap region.

Figure~\ref{figgalmap} shows the distribution of the $\sim710,000$ galaxies
with $I \le 23.5$ on the sky. In this figure, each galaxy is represented
as a point with intensity proportional to $10^{-0.4m}$ where $m$ is the isophotal
$I-$band magnitude. The galaxy distribution in this figure incorporates the
photometric and astrometric calibrations described above as well as the statistical
correction for object misclassification described in \S\ref{sgclass}.
Blank rectangular areas are the regions excluded due to the presence of bright stars.

\section{Cluster Detection}
\label{cluscat}

We use the matched filter algorithm developed by \citet{p96} to detect and
characterize cluster candidates in our survey. The algorithm employs
flux and radial filters to optimize the contrast of a distant cluster
projected on the sky. For details of the matched filter algorithm, and its
variants, readers are referred to 
\citet{p96}; \citet{sb98}; \citet{k99}; and \citet{k01}.
In addition to detecting clusters, this algorithm
provides a richness estimate, $\Lambda_{CL}$,
and an estimated redshift for each candidate cluster. $\Lambda_{CL}$ is
the effective number of $L^*$ galaxies in the cluster and within the
cutoff radius of the radial filter. 
We have modified the computation of $\Lambda_{CL}$
as proposed by \citet{sb98} to minimize redshift
dependent bias. 

Cluster detection is run on an image that is a convolution of the matched
filter, tuned to a specific redshift, with the galaxy catalog. 
Regions excluded from object detection (see \S\ref{galcat}) 
are also excluded from cluster detection. As in \citet{p96}, the matched filter
is tuned to 11 different redshifts between 0.2 and 1.2 in 0.1 intervals.
The estimated redshift is determined by choosing the redshift that maximizes
the matched filter signal for a given detection. Any intrinsic cluster parameters
(\eg $\Lambda_{CL}$, effective radius, etc.) are then determined for this redshift.
The peak of the flux filter in magnitude-space depends on the assumed
cluster galaxy luminosity function.  Table 1 details the
parameters used in our cluster detection process.
We apply evolutionary and k-corrections to the assumed characteristic magnitude, 
m$^*$, that are appropriate for a passively evolving elliptical galaxy. Specifically,
we assume $M^*(z) = M^*(z=0) - z$, which is consistent with the observations
of distant cluster ellipticals in \citet{pol} and \citet{fum}. The $z=0$ $M^*$ value
in Table 1 is derived from $R-$band observations of low redshift clusters and
transformed to the $I-$band using transformations in \citet{fg94}. 
 
Once cluster detection is complete, we minimize the problem of ``deblended" clusters
(\ie clusters detected multiple times due to substructure) by identifying 
cluster candidates whose effective radii overlap and whose estimated redshifts
lie within $\Delta z = 0.2$. For each overlapping pair, we exclude from the
final catalog the detection with the lower matched filter signal. 
Typically, this procedure reduces the size of the final cluster sample by $\sim5 - 10$\%.

\subsection{The Cluster Catalog} 

\begin{figure*}[t]
\leavevmode
\epsfxsize=6.5inch
\epsfbox{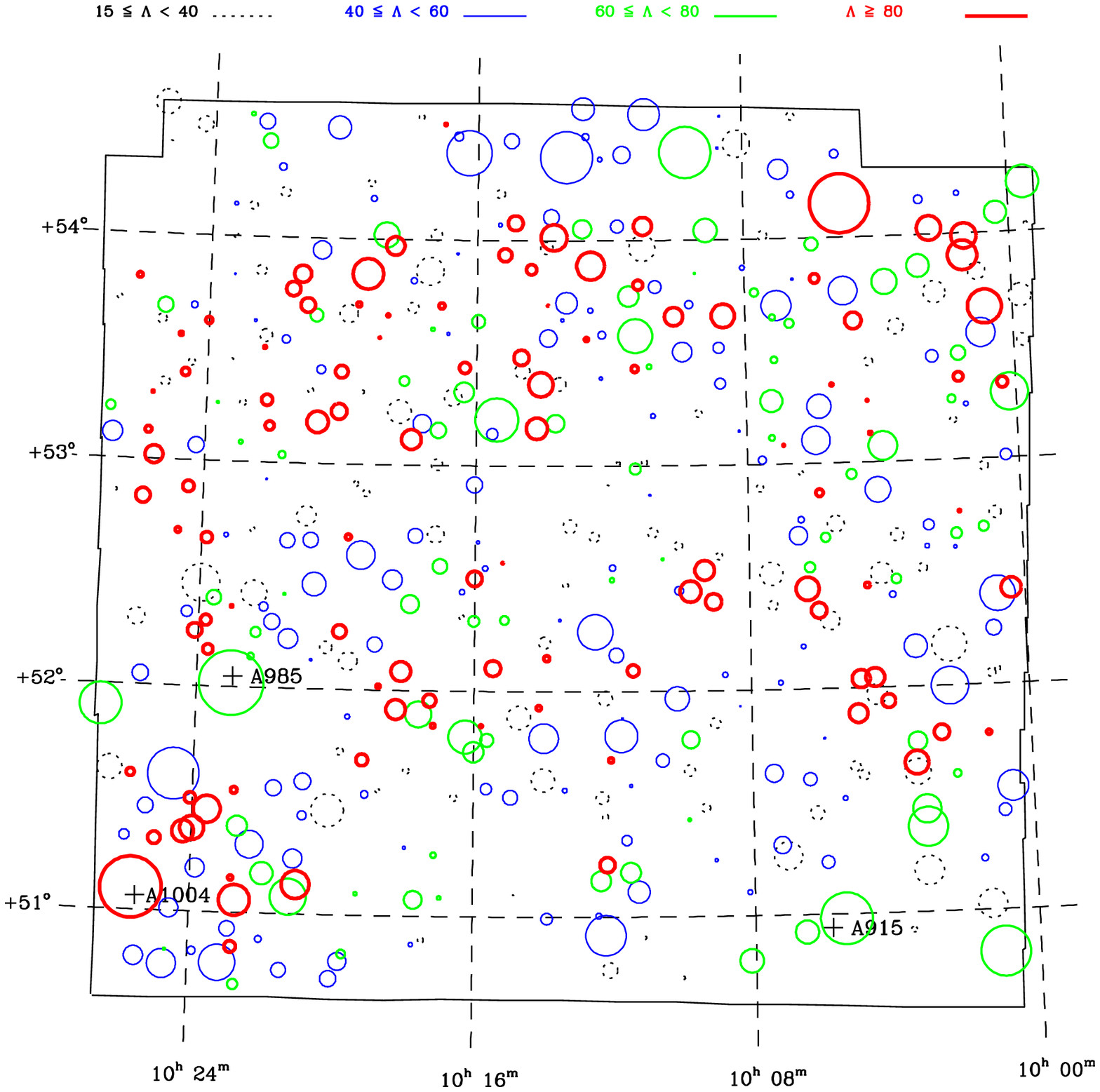}
\caption{The distribution of cluster candidates on the sky. Circles denote
the cluster position and their size
is directly proportional to the cluster's effective radius. Circle
line thickness is related to the $\Lambda_{CL}$ value. Dashed circles
have $\Lambda_{CL} < 40$, thin-lined circles have $40 \le \Lambda_{CL} < 60$,
medium-lined circles have $60 \le \Lambda_{CL} < 80$, and thick-lined
circles have $\Lambda_{CL} \ge 80$. The survey boundaries are shown for reference.
The locations of the 3 Abell clusters that lie within our survey are denoted
by crosses along with their Abell ID number.
}
\label{figclusmap}
\end{figure*}

Table 2 lists the 444 cluster candidates in the Deeprange survey with
a detection significance of $2\sigma$ or greater. As this
a single passband survey, clearly many of the $2\sigma$ detections
will be spurious. Visual inspection suggests that some higher $z$ candidates
($z \gsim 0.8$) with detection levels in the 2 -- 3$\sigma$ range are likely
to be real. We have thus opted for completeness over ``cleanliness" in publishing
this catalog. However, we fully characterize the false positive rate as a function
of detection level, $\Lambda_{CL}$, effective radius, and redshift in \S\ref{fprate}.
Users should carefully review that section before deciding what minimum detection
significance, effective area, and cluster richness they wish to use.
A sample of cluster candidates with a minimal number
of false positives can easily be constructed by appropriately filtering on these
parameters. 
Figure~\ref{figclusmap} shows the distribution of the cluster candidates on
the sky. The radius of each circle is directly proportional to the cluster effective
radius. The thickness and style of the circle line is determined by the $\Lambda_{CL}$
estimate of the cluster. 

A numeric cluster ID is given in column 1 of Table 2. The mean J2000 right ascension
(h m s) and declination (d m s) of each cluster candidate are given in columns 2 through 7.
The redshift estimate is given in column 8 and the $\Lambda_{CL}$ value is given
in column 9. Columns 10 through 13 contain four Abell-like richness parameters.
The first, $N_{A,0.25}$, is the number of galaxies within a 333$h_{75}^{-1}$ kpc
radius that lie within $m_3$ and $m_3 + 2$, where $m_3$ is the statistically selected
3rd brightest cluster galaxy in this same radius. A statistical background subtraction,
derived from galaxy counts in survey regions with no detected cluster candidates,
is applied to all the Abell-like richness estimates. The $m_3$ value is given in
column 16 of the table. The next Abell-like richness estimate
is $N_{e,0.25}$ and is computed over the magnitude range $m_{Eff}$ to $m_{Eff}+2$,
where $m_{Eff}$ is the apparent magnitude at which the cluster is maximally enhanced
over the background field. The $m_{Eff}$ is derived by the matched filter algorithm
and is given in column 17. The final two Abell-like richness estimates are
$N_{A,0.5}$ and $N_{e,0.5}$, which are similar to those above except that the computations
are done out to a radius of 666$h_{75}^{-1}$ kpc ($m_3$, however, is the value
derived for the central 333$h_{75}^{-1}$ kpc region). The significance of each detection
(in $\sigma$ units) is listed in column 14 and the effective radius (in arcseconds)
is in column 15. The effective radius is $\sqrt{A_{MF}/\pi}$, where $A_{MF}$
is the area of the detected signal in the match filtered image. 
Lastly, column 18 contains a visually
assigned quality flag for all candidates with $\Lambda_{CL} \ge 50$. This flag
is assigned by M.P. and is his attempt to assess the reality of each detection.
A value of 3 indicates the cluster candidate is almost certainly real -- a 
substantial galaxy enhancement is seen and the galaxies exhibit a pronounced central
concentration similar to spectroscopically confirmed clusters at similar redshifts. A value
of 2 indicates that the cluster candidate is likely to be real but may also be
a compact group. A value of 1 indicates that the detection does not appear to
have a strong central concentration of galaxies and may very well be spurious,
a poor group, or deblended substructure from a nearby rich cluster. 
Use of the visual quality flag is meant for non-statistical analyses only.

\subsection{The Selection Function}
\label{selfunc}

Meaningful astrophysical constraints can only be extracted from this cluster
catalog when the selection function and false positive rate are well
measured and a good calibration of the richness 
parameter, $\Lambda_{CL}$, is available. 
Implanting simulated clusters into the actual galaxy catalog
is the most reliable method for determining the selection function and richness
calibration. We have simulated over 17,000 clusters to obtain the required
precision in our selection function and richness calibration. 
Specifically, we generate clusters spanning the ranges $10 \le \Lambda_{CL} \le 1000$,
$0.2 \le z \le 1.2$. At each redshift and richness, we also create clusters
dominated by an elliptical galaxy population and clusters dominated by an Sbc
galaxy population. By this we mean that the evolutionary and 
k-corrections applied are appropriate
either for a passively evolving elliptical or Sbc dominated galaxy population. 
Clusters at $z \sim 1$ can exhibit spiral fractions that are 3 -- 5 times
larger than that seen in nearby clusters (\eg \citealt{vd2001, lubhst, lubpcomm}).
Finally, we also create clusters with surface density profile
slopes of $-1.0, -1.4,$ and $-2.0$. The observed range in
cluster profile slopes is typically $-1.6 \le \gamma \le -1$ 
(\eg \citealt{p88, tf95, lp96, sq96}). 
We have done the same for clusters with
richness corresponding to Abell richness classes 0 through 4. These simulated
cluster galaxies are inserted into the actual galaxy catalog in locations not
occupied by real detected clusters. The cluster detection software is then
run (using the identical detection parameters as in the real catalog)
and we record the resulting $\Lambda_{CL}$, detection significance, 
and estimated redshift as well as the false negative rate.
The selection function can then be computed as a function of all of these
key parameters. 

\epsfxsize=\hsize
\centerline{\epsfbox{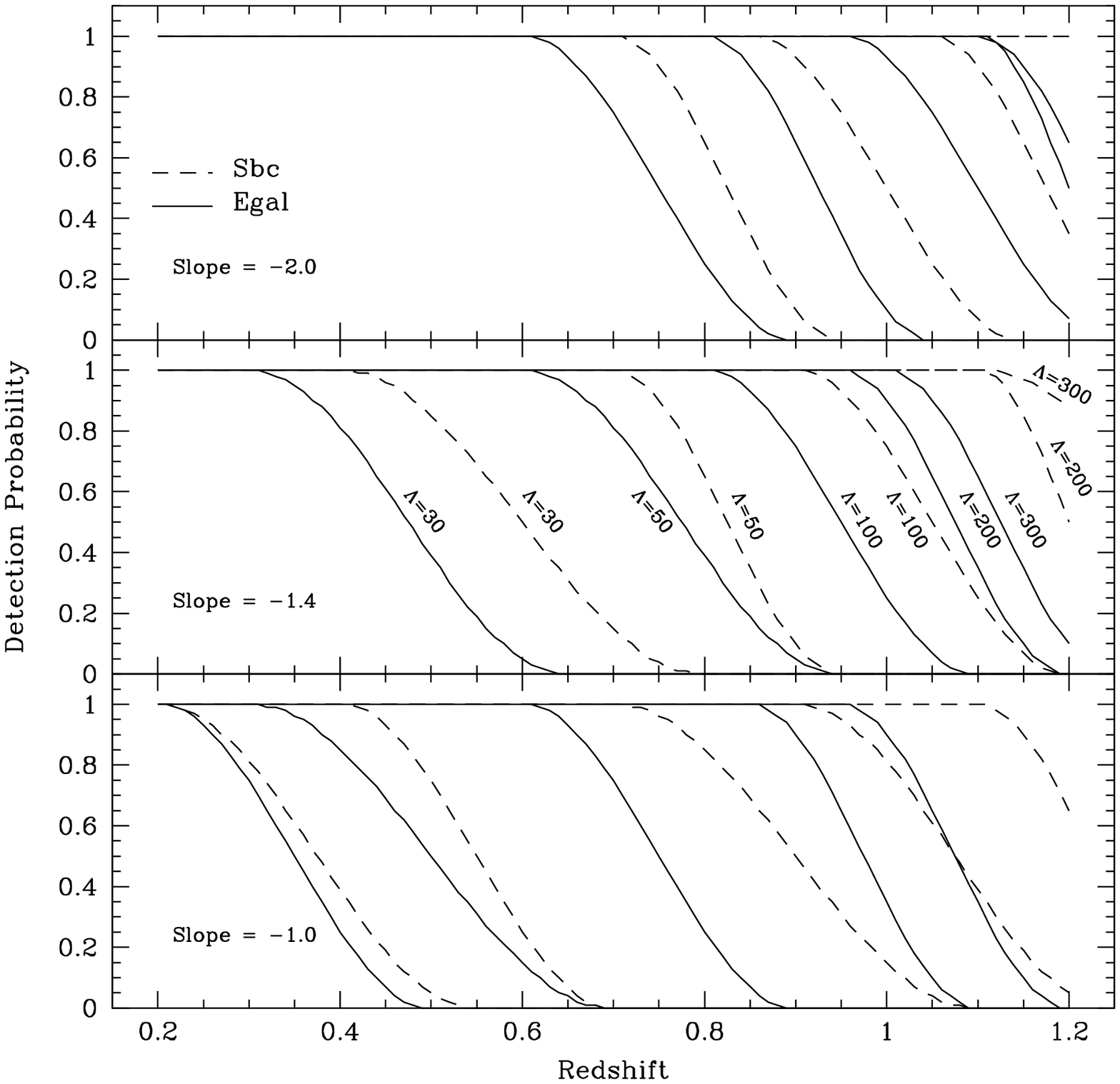}}
\figcaption{\footnotesize
The cluster detection probability as a function of redshift,
$\Lambda_{CL}$, cluster surface density profile slope, and
galaxy spectral type. Results are explicitly shown for our
$\Lambda_{CL} = 30, 50, 100, 200,$ and $300$ simulations. The
solid curves are for the simulations using an elliptical dominated
galaxy population. The dashed curves are for an Sbc dominated galaxy
population. The $\Lambda_{CL}$ values are explicitly labeled in
the middle (slope$=-1.4$) panel. The order of the curves in the
other panels is similar -- for lower $\Lambda_{CL}$ clusters,
detection probabilities fall off at lower redshifts.
\label{figselfunc}}
\vspace{\baselineskip}
Figure~\ref{figselfunc} shows the cluster detection probability
as a function of $\Lambda_{CL}$, profile slope, galaxy population, and redshift
for all detections with significance $\ge2\sigma$. The detection efficiency
decreases with increasing redshift, decreasing richness, and
decreasing star formation rate. Furthermore, clusters that are more centrally
concentrated are detectable to higher redshifts than their more diffuse counterparts.
The detection probability as a function
of redshift for a given richness, galaxy population, and profile slope
typically declines from 1 to 0 over a range $\Delta z \sim 0.2$.
We use the selection functions computed above in all computations of cluster
space density and in the estimation of the $\Lambda_{CL}$ distribution function.

\subsection{Richness Calibration}
\label{richcal}

There are essentially two methods to calibrate the richness parameter
$\Lambda_{CL}$ that enable us to compare our derived space densities
with those from the Abell cluster catalog. The first
method takes advantage of the extensive simulations performed to compute
the selection function. These simulations provide two key results: the
accuracy with which we recover the $\Lambda_{CL}$ value of a given cluster 
and the relationship between $\Lambda_{CL}$ and the Abell richness count.
The relationship between $\Lambda_{CL}$ and the Abell richness count can
also be obtained empirically by processing known Abell clusters through
our algorithm. For this empirical calibration to be most reliable, the
imaging data of the Abell clusters should be identical to that in this survey.
This would be time-intensive to perform for a large sample of Abell clusters.
Fortunately, there are 3 Abell clusters in the survey boundaries 
(A915, A985, A1004) and these provide at least a rough reality check 
on the more precise calibrations provided by the simulations.
\vspace{\baselineskip}
\epsfxsize=\hsize
\centerline{\epsfbox{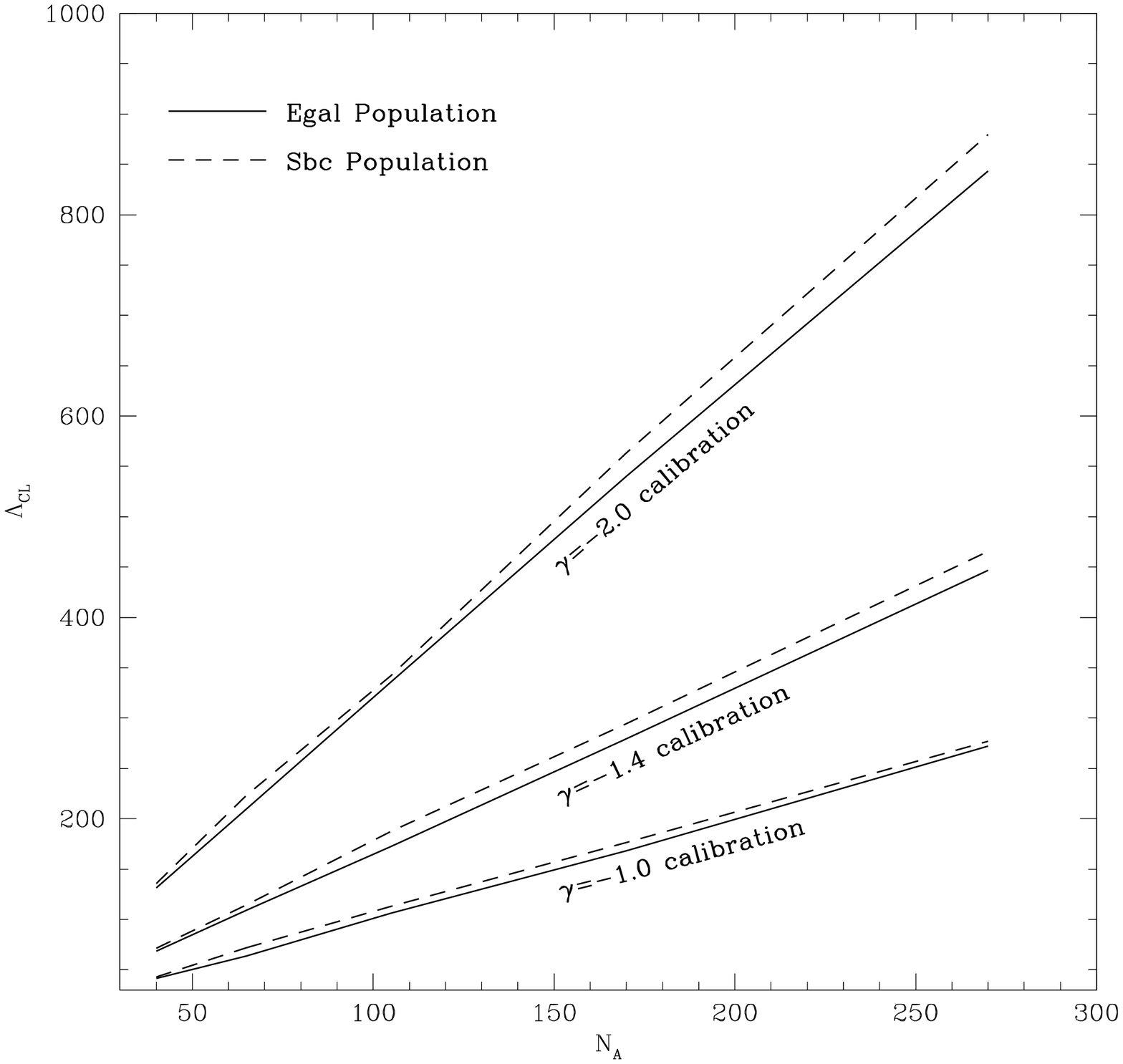}}
\figcaption{\footnotesize
The relationship between $\Lambda_{CL}$ and the Abell richness
count, $N_A$, as estimated from simulated cluster data run through the
matched filter algorithm. The simulations span the Abell richness class
range $0 \le {\rm RC} \le 4$ and are constructed using one of three
different surface density profile slopes: $-1, -1.4, {\rm or} -2$.
\label{figrcal}}
\vspace{\baselineskip}

We use the same simulation techniques used to compute the cluster selection
function to create, insert, and measure the characteristics of clusters
with known Abell richness classes. As in the selection function computations,
we generate clusters with surface density profile slopes of $-1, -1.4,$ and
$-2$ and with either Sbc or Elliptical dominated galaxy populations. The calibration
is only weakly dependent on the galaxy type but is strongly dependent on
the cluster profile slope. The slope dependence is largely a consequence of
the use of a radial filter that weights centrally located cluster 
galaxies more than those at larger radii -- a system with a steeper profile
has relatively more galaxies located near the cluster center than an equally
massive cluster with a flatter profile. 
Figure~\ref{figrcal} shows the derived $\Lambda_{CL}$
values as a function of the Abell richness count, $N_{A}$, for simulated clusters
with the 3 different profile slope values.
The relationship between $\Lambda_{CL}$ and $N_{A}$ is 
linear for a given surface density profile. Taking the average between the
Egal and Sbc calibrations, we find that a reasonable representation of the
calibration between $\Lambda_{CL}$ and $N_{A}$ is
\begin{equation}
\Lambda_{CL} \approx f(\gamma) N_A\ {\rm where}\ 
f(\gamma) = 0.46 + 0.26\gamma + 0.81\gamma^2
\end{equation}
and where $\gamma$ is the slope of the cluster surface density profile.
Figure~\ref{figabrcal} shows a zoomed view of these relations with
the actual data from the 3 Abell clusters superposed. The Abell cluster
results lie in between the $\gamma=-1$ and $\gamma=-1.4$ calibration.
The best-fit line (with a zero-valued intercept) to
the cluster data in Figure~\ref{figabrcal} is 
\begin{equation}
\Lambda_{CL} = 1.24 N_A
\end{equation}
corresponding to an average profile slope of $\gamma \sim -1.16$ (equation 5). 
This average is consistent with the actual slopes we measure for these 3 clusters:
$-1.23\pm0.04$ (A915), $-1.00\pm0.01$ (A985), and $-1.19\pm0.02$ (A1004).
We thus adopt equation 6 as our nominal relationship between $\Lambda_{CL}$ and $N_A$.
The Abell clusters also provide a calibration of the Abell-like
richness parameter, $N_{A,0.5}$, provided in Table 2. We find the
relationship is $N_{A,0.5} \approx 0.44 N_A$.
 
\epsfxsize=\hsize
\centerline{\epsfbox{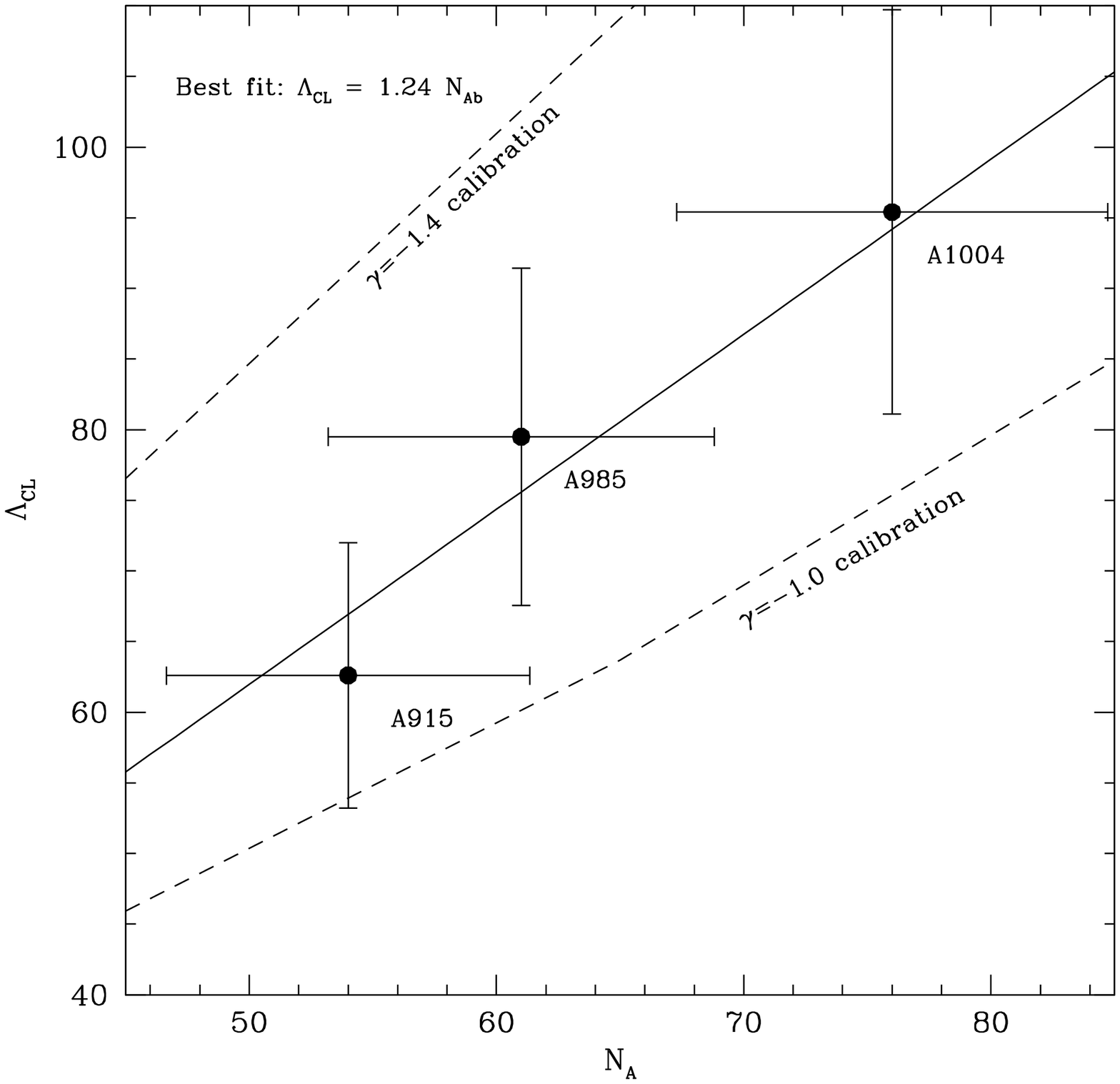}}
\figcaption{\footnotesize
The empirical relationship between $\Lambda_{CL}$ and $N_A$
for Abell clusters A915, A985, and A1004. These 3 clusters lie wholly
within the survey boundaries. The dashed lines show the same relations
depicted in Figure~\ref{figrcal}. The solid line is the best-fit line
with a zero-valued intercept. Errors in $\Lambda_{CL}$ are $\pm15$\%, as
derived from our simulations. Errors in $N_A$ are just $\sqrt{N_A}$.
\label{figabrcal}}
\vspace{\baselineskip}
Out to $z_{est} = 0.9$, we find that
the typical error in $\Lambda_{CL}$ is $\pm15$\%. This is measured by comparing
the known $\Lambda_{CL}$ value from each simulated cluster
at a given richness and redshift with the value computed by the matched filter
algorithm. The $\Lambda_{CL}$ uncertainties are insensitive to redshift
in the range $0.2 \le z_{est} \le 0.9$. Beyond $z_{est} = 0.9$,
the corrections used to compensate for the truncation of the flux filter
by the survey flux limit become quite sizeable and the errors in $\Lambda_{CL}$
can reach 40\% (see section 4.1 in \citet{p96} for details).
Our $\Lambda_{CL}$ values are thus most reliable for
those candidates with $z_{est} \lsim 0.9$.
 
\subsection{False Positive Rates}
\label{fprate}

There are two basic sources for false positive detections in this survey:
random fluctuations of field galaxies and chance
alignments of poor groups of galaxies. The false
positive rate for random fluctuations in the galaxy distribution can be
assessed by creating simulations that contain no clusters but that have
an angular two point correlation function similar to the galaxies in the
real universe. We generate such simulations using a Rayleigh-Levy galaxy
pair separation distribution: $P(>\theta) = (\theta /\theta_o)^{-d}$ if 
$\theta \ge \theta_o$ and $P(>\theta) = 1$ if $\theta < \theta_o$. To construct
the clustered galaxy distribution, we randomly select a starting position within
the survey boundaries and then generate galaxy positions about this center
using the above separation distribution. We allow up to 7 galaxies to be so
distributed about a given center. The results below are not very sensitive to
this choice so long as the number is more than 2 and less than 15. 
A new center is then randomly chosen and the
process repeated until the number of galaxies matches the observed number.
The values of $\theta_o,\ d$ are chosen to match the observed $\omega(\theta)$
derived from this survey as best as possible \citep{dr1}. We then run the matched
filter algorithm on these clusterless fields using the identical detection
parameters as in the real survey. Figure~\ref{figfpr} shows the resulting
\vspace{\baselineskip}
\epsfxsize=\hsize
\centerline{\epsfbox{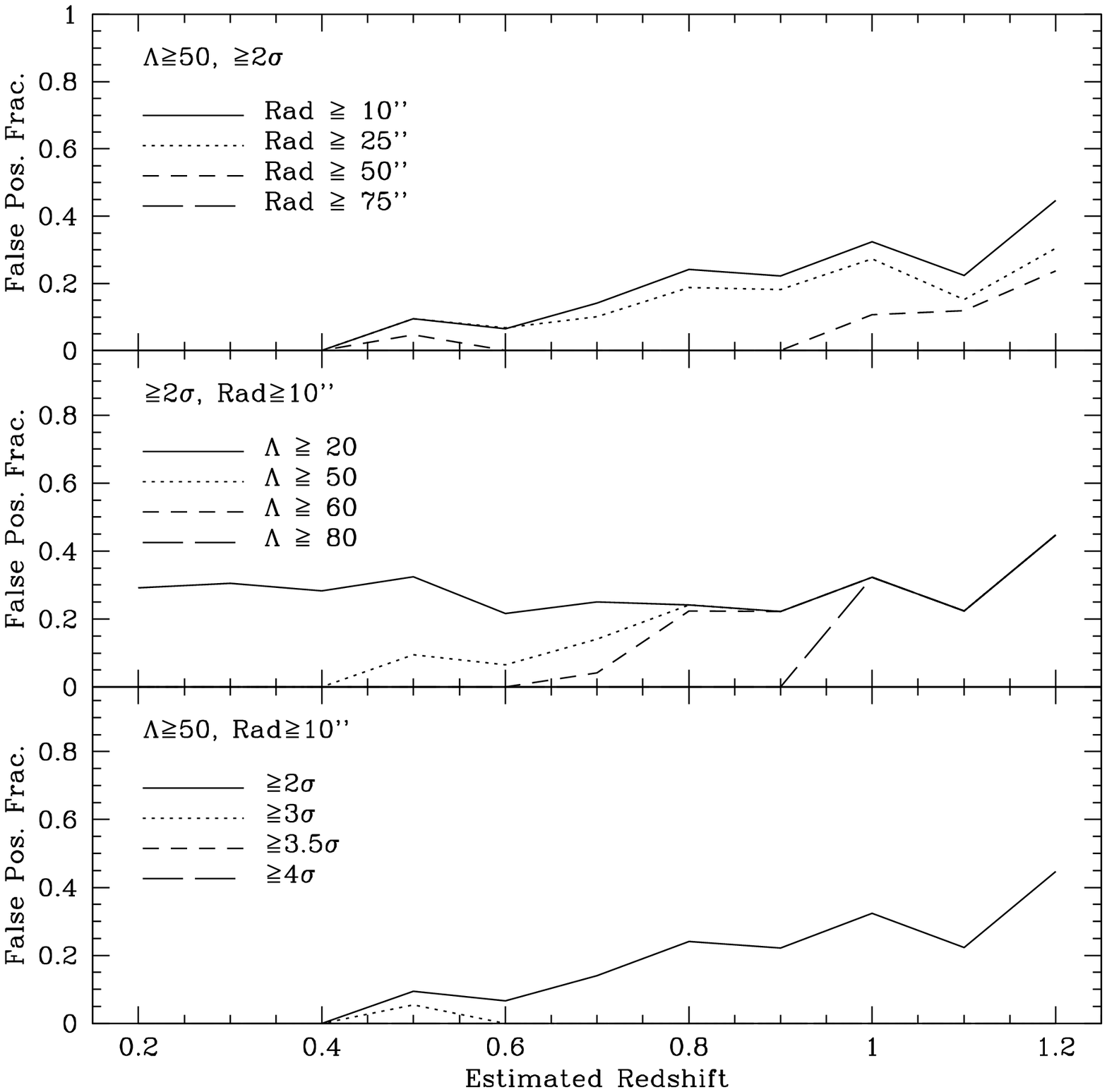}}
\figcaption{\footnotesize
The fraction of detections that could be spurious as a function
of redshift, richness, detection significance, and effective radius. These
results are obtained by running the matched filter algorithm on simulated fields with
a non-zero two point correlation function but with no clusters. The
false positive fraction value is computed by dividing the surface
density of detections in these simulations by the observed
surface density of cluster candidates in the actual survey.
\label{figfpr}}
\vspace{\baselineskip}
\noindent false positive fractions as functions of estimated redshift, $\Lambda_{CL}$,
detection significance, and effective radius. 
The false positive fraction is the number of detections in the clustered
simulation (per square degree) divided by the observed number of cluster
candidates (per square degree) found in the actual survey.
These simulations predict that
a sample of cluster candidates with $\Lambda_{CL} \ge 20$, $\sigma \ge 2$,
and effective radii $\ge 10''$ will suffer a false positive rate between
10\% to 40\% depending on the redshift. The vast majority of these spurious
detections lie at the low end of the $\Lambda_{CL}$ distribution and tend
to also be small in size. Table 3 shows how the number of clusters in the
survey varies with the minimum $\Lambda_{CL}$, $\sigma$, and effective radius.
Users who wish to work with a catalog that contains a minimum number of
spurious detections should use candidates with $\Lambda_{CL} \ge 60$,
$\sigma > 3$, and effective radii $\ge 60''$. Such a sample should have
few spurious detections due to large positive fluctuations in the field galaxy
distribution.  Note, however, that such selection criteria will 
significantly reduce the number of high redshift candidates in the sample.

Approximately 70 to 80\% of the cluster candidates with 
$z_{est} \lsim 0.7$ are most likely physically real galaxy associations. The false positive
rate for candidates with $z_{est} > 0.7$ will be assumed to be $\sim 30\ - 40$\%
for the purposes of computing true cluster space density estimates.
We note that the use of simulations with non-zero angular two point correlation functions
is essential in accurately estimating false positive rates. Simulations with purely
random galaxy distributions yield false positive rates that are an order of magnitude
smaller than that observed in the clustered simulations and in the actual data.
\vspace{\baselineskip}
\epsfxsize=\hsize
\centerline{\epsfbox{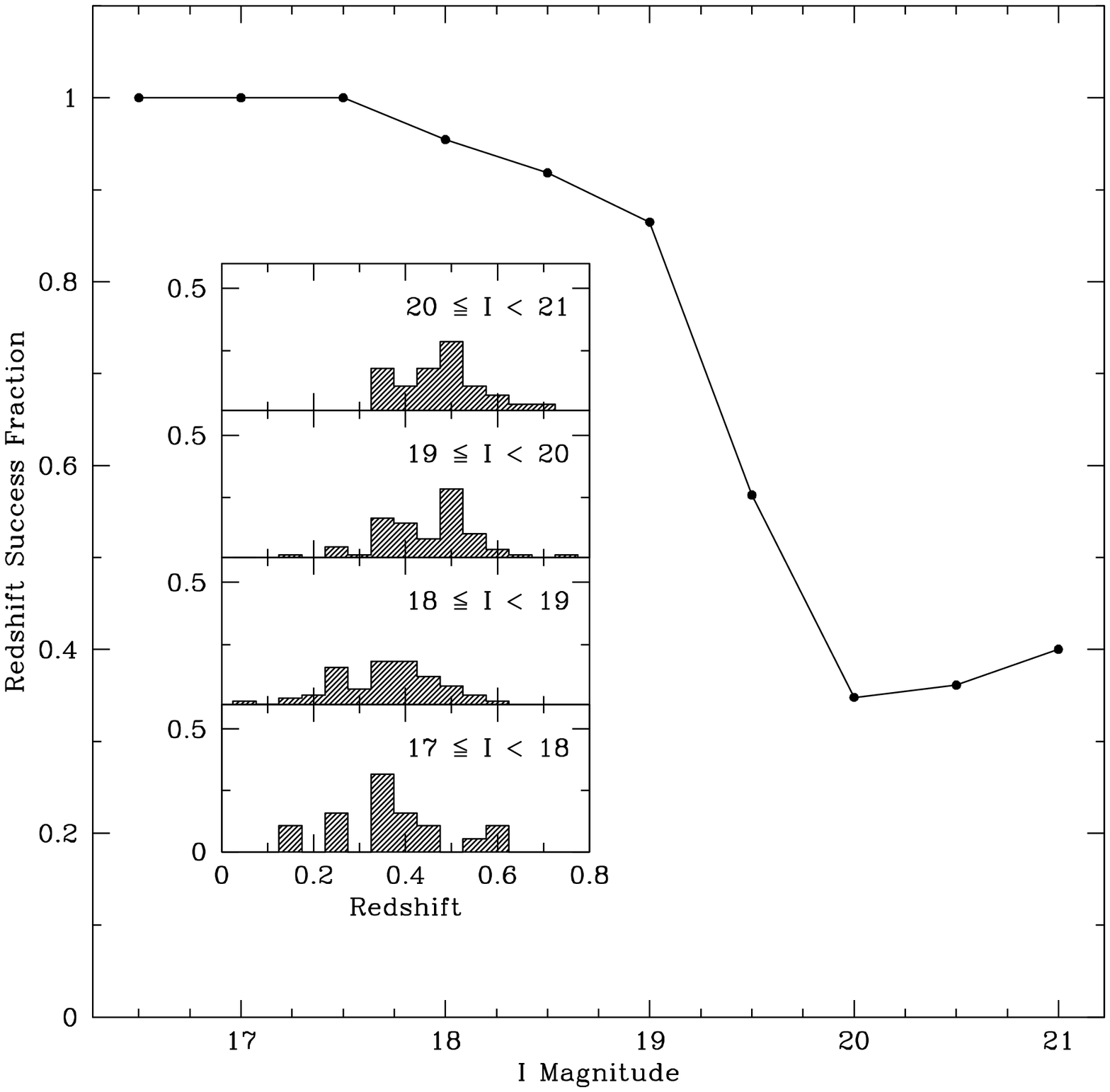}}
\figcaption{\footnotesize
The fraction of galaxies for which redshifts were
successfully measured as a function of the isophotal $I-$band
magnitude. Inserts show the redshift histograms in four
magnitude bins.
\label{figzsuc}}
\vspace{\baselineskip}

\section{Spectroscopic Follow-up}
\label{specfu}

Assessing the occurrence of spurious detections due to the
chance radial alignment of physically unrelated groups of galaxies (or
edge-on sheets of galaxies) is best addressed by spectroscopic follow-up
of a well selected subset. Indeed, a spectroscopic survey enables one
to measure the overall spurious detection rate due to both random fluctuations
and chance alignments. Such a follow-up survey was begun in December 1998
and we find that up to $z = 0.6$ the false positive
rate is $\sim25$\%. This rate includes all spurious detections regardless of their
nature. It is in excellent agreement with that predicted by the simulations above.
The scientific motivation for this spectroscopic survey is 
to measure the 3D cluster-cluster correlation function at $z \sim 0.5$. 
To achieve this goal, we set out to obtain
spectra for galaxies in the 141 clusters with $0.3 \le z_{est} \le 0.7$ and
$\Lambda_{CL} \ge 50$. As of March 2001, we have observed 36 clusters.

We have had a total of 6 spectroscopic observing runs on the KPNO 4m
telescope.  The first 4 runs (Dec 1998, Feb 1999, Jan 2000, Mar 2000) used the
Cryocam spectrograph. The Cryocam suffered a mechanical failure in March
2000 and our subsequent observations (Jan 2001, Mar 2001) were performed
with the less efficient RC spectrograph. Both spectrographs were used with the BL-181
grating to provide low resolution (7 -- 10\AA) spectra over the range 4200 -- 9500\AA.
We visually selected likely cluster members\footnote{Highest priority was
given to galaxies that fell within $\pm0.5$ mag of the effective cluster
magnitude derived by the matched filter algorithm.} within 2.5 arcminutes 
from the cluster center (the Cryocam field of view is 5 arcminutes) 
using our $I-$band images. KPNO provided software that was then used
to create slit mask templates. We produced one mask per cluster 
and the masks typically contained between 8 to 12 slits (a slit width
of 2 arcseconds was used).   

We always obtained at least 2 exposures per cluster and, more typically,
3 to 4 exposures. Depending on the estimated cluster redshift and the
observing conditions, the combined exposure times range from 3600 seconds 
(for a $z_{est} = 0.3$ cluster observed with Cryocam 
in clear weather with good seeing) to 9500 seconds (for a $z_{est} = 0.5$
cluster observed with RCSpec in less than ideal conditions). 
The accuracy of our astrometry proved sufficient to assure successful
centering of galaxy targets in their slits but we note that the use
of check and setup stars was critical in achieving this good alignment.
The KPNO-4m$+$Cryocam/RCSpec are well suited to obtaining redshifts 
for systems with $z_{est} < 0.6$ but the availability of an 
imaging spectrograph at NOAO would significantly increase the efficiency of 
obtaining such observations. The relatively modest productivity of our 
spectroscopic program (36 clusters over 2.3 years) is largely due to
weather related problems. None the less, we did succeed in completing
the observations for all clusters with $0.3 \le z_{est} < 0.5$.
Acquiring redshifts efficiently for the
more distant component of our sample (81 clusters with $z_{est} \ge 0.6$) 
will require a 6 - 8m class telescope.  

The spectroscopic data were extracted and calibrated using standard IRAF
longslit packages. We did observe small sub-pixel shifts in the sky line
positions between subsequent Cryocam exposures of the same target. This
required that the multiple exposures be co-added {\it after} extraction
and wavelength calibration. The RC Spectrograph data were more stable and
thus could be co-added prior to spectral extraction and calibration.

\begin{figure*}[t]
\leavevmode
\epsfxsize=6.5inch
\epsfbox{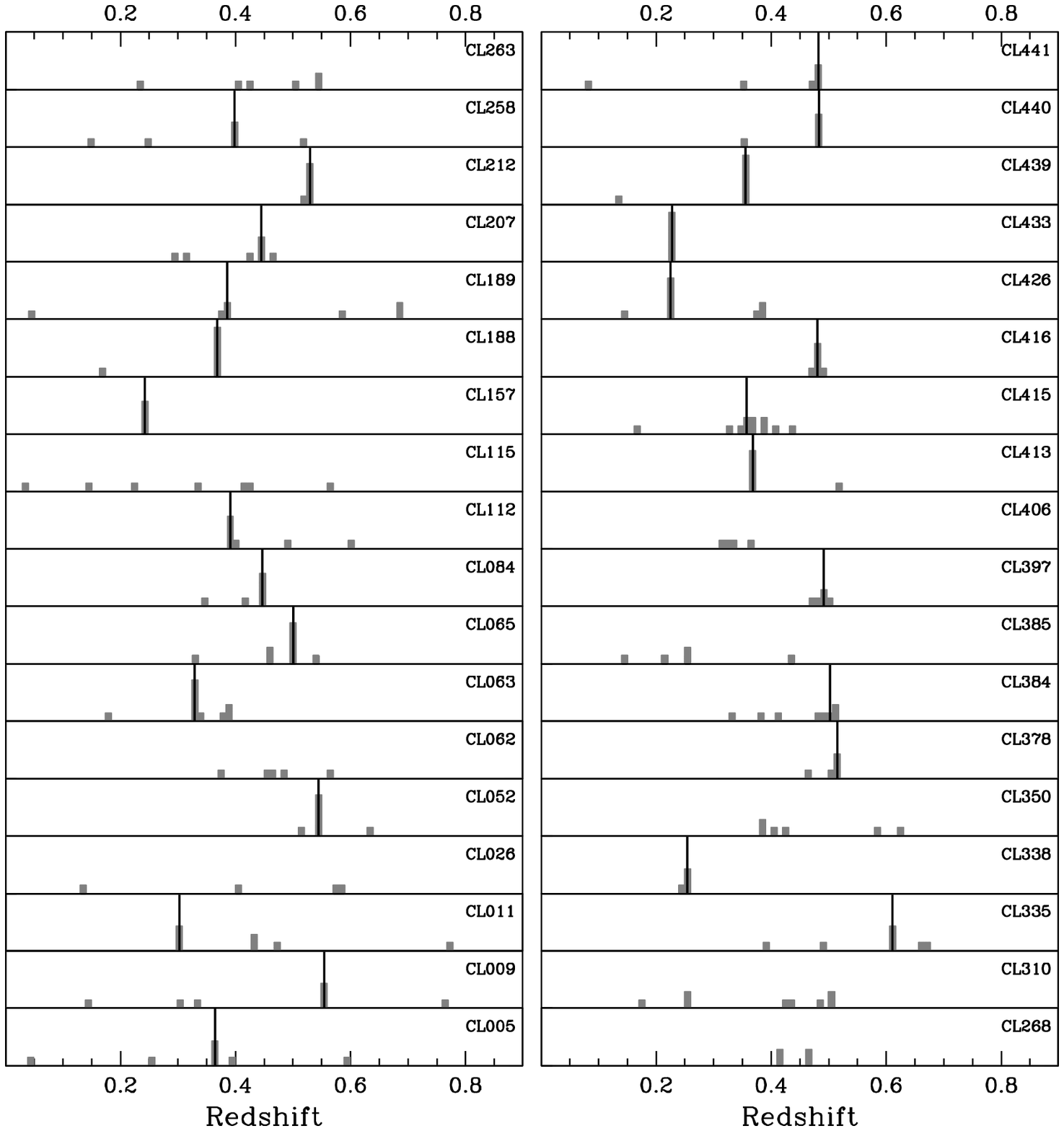}
\caption{
The redshift distributions for the 36 clusters candidates
observed in our spectroscopic follow-up survey. Of the 36, 9 are
probably spurious or superpositions of poor groups. The vertical
lines indicate the adopted mean cluster redshift, when a reliable mean
can be determined.}
\label{figzhist}
\end{figure*}

\subsection{Galaxy and Cluster Redshift Determination}

Galaxy redshifts were derived using the IRAF package {\it xcsao}. We
used 8 independent high S/N templates of elliptical galaxies to 
perform the cross-correlations. Regions around prominent night sky
lines (Hg, NaD, OI) and strong atmospheric OH absorption bands 
were excluded from the fitting procedure. 
A galaxy redshift was computed by averaging the results for those templates
that lie within 250 km $s^{-1}$ of the mode. For a high S/N spectrum,
all 8 templates are in agreement. If fewer than 4 templates are found
to agree, the spectrum is deemed unusable for redshift determination.
Of the 352 spectra that were obtained, we were able to derive reliable 
redshifts for 230. The remaining spectra had insufficient S/N.
Figure~\ref{figzsuc} shows the redshift measurement success rate
as a function of $I-$band magnitude along with redshift histograms
as a function of magnitude. 
All spectra were visually inspected as well to assure
the resultant redshift was sensible. 
Approximately 10\% of all the spectra exhibited emission features and redshifts
for these galaxies were determined using the IRAF package {\it rvidlines}.
Table 4 gives the positions, $I-$band magnitudes, redshifts and redshift 
errors for all the observed objects.

Cluster redshifts are derived by looking for ``clumps" in redshift that
are not wider than $3000$ km $s^{-1}$ in the rest frame. If at least 3 galaxies can
be identified in such a clump, we average their redshifts to compute
a mean value for the cluster. The results are listed in Table 5 and displayed
in Figure~\ref{figzhist}.
Of the 36 clusters observed, two (CL415 and CL426)
appear to be superpositions of two
systems. In both cases, however, one of the spikes in redshift contains
5 concordant redshifts indicating the presence of a real cluster. For 6 of
the 36 clusters (CL026, CL062, CL115, CL263, CL268, CL385), 
no reliable redshift could be derived because no clump
of 3 or more concordant redshifts was identified or because the S/N of the 
spectra for most of the objects was too low. Finally, for 3 of the clusters
(CL310, CL350, CL406) a ``clump" of 3 redshifts exists but is just larger 
than $\sim3000$ km $s^{-1}$
in width. If we assume all 9 of these cluster candidates are spurious, then our
matched filter algorithm is finding real physical systems $\sim75$\% of the
time (at least out to $z \sim 0.6$), 
in accord with other studies \citep{h97} and with simulations
done in this paper and by \citet{p96}.
\vspace{\baselineskip}
\epsfxsize=\hsize
\centerline{\epsfbox{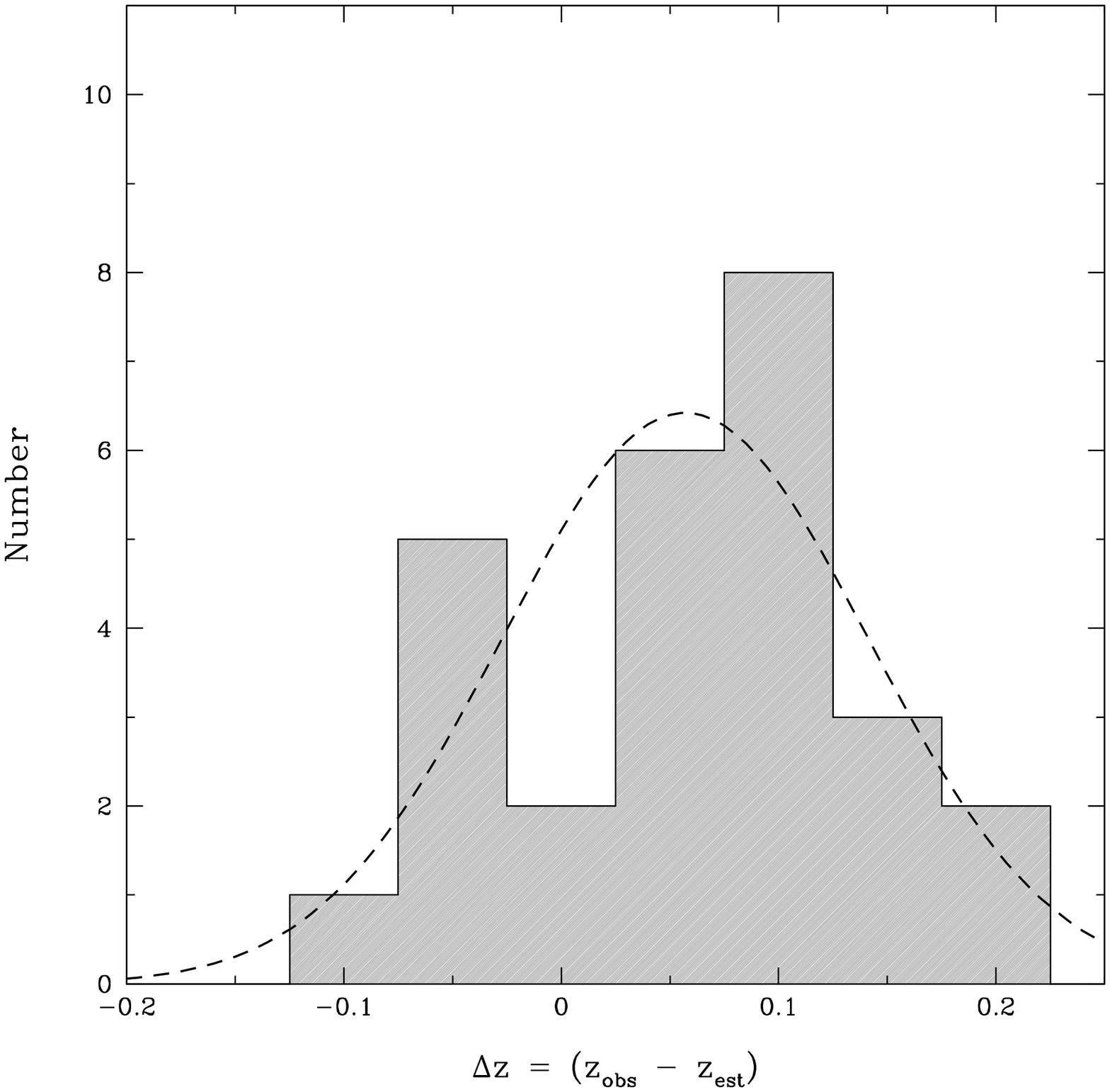}}
\figcaption{\footnotesize
The distribution of differences between the observed
spectroscopically derived cluster redshift and the redshift
estimate derived from the matched filter algorithm. The data
are derived from the 27 clusters with 3 or more concordant redshifts
within $\pm$1500 km s$^{-1}$. The best
fit Gaussian representation of the distribution
($\overline{\Delta z} = +0.057,\ \sigma_z = 0.084$) is also shown.
\label{figzcomp}}
 
\subsection{Comparison with Estimated Redshifts}
\label{zest}

The accuracy of the matched filter redshift estimates depends largely
on the accuracy of our cluster luminosity function model. 
Figure~\ref{figzcomp} shows the redshift difference histogram
($\Delta z = z_{obs} - z_{est}$) for the 27 clusters for which reliable mean
redshifts could be measured. The mean offset ($\overline{\Delta z}$) 
is 0.057 with an rms scatter ($\sigma_z$) of 0.084. 
These statistics are largely independent of 
the redshift estimate over the range $0.3 \le z_{est} \le 0.5$. 
The relatively good agreement between the observed and estimated 
redshifts indicates that
our cluster LF model is reasonable and further suggests that the
characteristic cluster galaxy magnitude is a reasonably reliable
($\pm$20\%) distance indicator. The slight positive offset of the
$\Delta z$ distribution can possibly be attributed to a small 
underestimate in the assumed characteristic luminosity, $L^*$.
However, the offset is smaller than the 0.1 interval used
to estimate redshifts.
Spectroscopic follow-up of clusters in the Palomar Distant Cluster survey
\citep{cops} suggests that the scatter observed above increases a bit
at higher redshifts ($z \lsim 0.8$): for 7 PDCS clusters
with $0.5 \le z_{est} \le 0.8$ $\sigma_z = 0.16$. This is consistent
with the growth in the expected systematic errors with redshift
(see \S\ref{spaceden})
but is still small enough that reliable space density estimates can
be established.

\section{Results}
\label{results}

The abundance of galaxy clusters as a function of redshift and mass
is a critical constraint on cosmological models and models for cluster
formation and evolution. With an accurate construction of the selection
functions, $\Lambda_{CL}$ calibration, and false positive rates complete,
we can now use our observations to constrain the observed space density
of optically selected clusters up to $z \sim 1$.
\vspace{\baselineskip}
\epsfxsize=\hsize
\centerline{\epsfbox{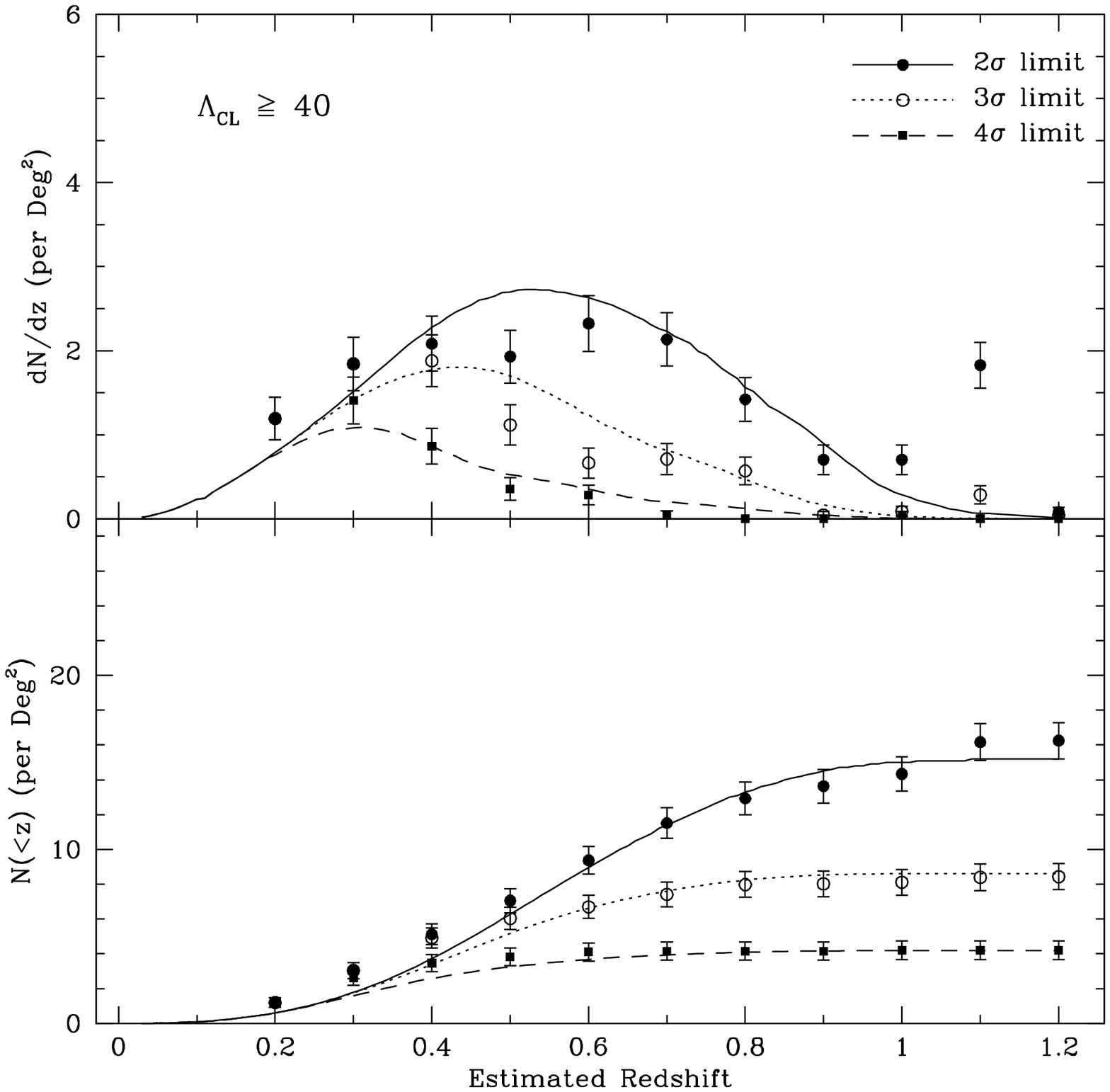}}
\figcaption{\footnotesize
The differential and cumulative surface densities
of clusters as functions of estimated
redshift and detection significance for systems with $\Lambda_{CL} \ge 40$.
The surface densities have been corrected for false positive detections.
The curves show the predicted relations for a constant comoving space
density of clusters multiplied by the relevant selection functions.
The best-fit constant comoving density (for $\Omega_m = 0.2, \Lambda = 0$)
that matches the surface density observations
is $(1.57 \pm 0.15) \times 10^{-5}\ h_{\rm 75}^{3}$ Mpc$^{-3}$, $\sim1.3$ times
the local space density of RC $\ge 0$ Abell clusters.
\label{figsurfden}}

\subsection{The Surface Density of Clusters vs Redshift}
\label{surfden}

The differential and cumulative surface density of $\Lambda_{CL} \ge 40$
cluster candidates as functions of redshift and detection threshold are shown in
Figure~\ref{figsurfden}. The measured counts have been corrected
for the expected false positive rates. The continuous curves in this figure
are the predictions for a constant comoving space density of clusters multiplied
by the appropriate selection functions. We also assume that the
relative abundances of clusters of different richness classes remain
independent of redshift.
We adopt \citet{dp02} estimates of the space density of Abell
clusters: $(1.17 \pm 0.12) \times 10^{-5}h_{75}^{3}$ Mpc$^{-3}$ for RC $\ge 0$
and $(3.80 \pm 0.72) \times 10^{-6}h_{75}^{3}$ Mpc$^{-3}$ for RC $\ge 1$.
For RC $\ge 2$, we assembled a compilation of redshifts, including data from
the MX Survey \citep{sl98} and the 2dF Survey \citep{dp02},
and derived our own space density estimate
of $9.16 \times 10^{-7}h_{75}^{3}$ Mpc$^{-3}$ based on a sample of 60 Abell clusters
with observed redshifts $z \le 0.1$ and $\vert b\vert \ge 40^{\circ}$.
The good agreement between the observed cluster surface densities and those
predicted by our convolution of the selection functions with a constant comoving
space density supports the hypothesis 
that there is little evolution in the comoving space density of clusters over 
the range $0 < z < 0.8$.
The constant comoving space density that best fits the surface density data is
$(1.57 \pm 0.15) \times 10^{-5}h_{75}^{3}$ Mpc$^{-3}$, about $1.3$ times
the mean space density of local RC $\ge 0$ Abell clusters. 
From equation 6, we find that
$\Lambda_{CL} > 40$ is approximately equivalent to $N_A > 30$, which is
the Abell RC = 0 threshold. 
The Deeprange cluster surface densities are consistent with those found
in other optical surveys using the matched filter (\eg PDCS \citep{p96} and
EIS \citep{eis}). The EIS and PDCS both find cluster surface densities 
in the range $5 - 10$ deg$^{-2}$ for $z_{est} \le 0.7$ and $\Lambda_{CL} \ge 40$.
The ratio of the space density of distant clusters
to that for the Abell catalog is, however, significantly
less than that reported by \citet{p96} for 
the Palomar Distant Cluster Survey. We believe the difference between the current
results and those from the PDCS is a result of the
more accurate $\Lambda_{CL}$ -- Abell richness calibration used here,
improved measurements of the local cluster space density (\citealt{p96}
used rather out-of-date estimates of the Abell cluster densities based on
results in \citealt{b79}), 
and the significantly larger contiguous survey area, 
which enables a more reliable measurement of the mean density level.
As we will demonstrate below, the somewhat larger space density of poor
clusters in our survey relative to the Abell catalog is attributed
to incompleteness at the low richness end of the Abell catalog.

The good agreement between the observed and predicted cumulative cluster surface density
also implies that we can reliably divide out the signature of the selection function
to obtain a constraint on the true cumulative cluster surface density.
The results of this computation are provided in Table 6. 
Table 6 also lists the effective survey area as a function of redshift.
The survey area is a weak function of redshift because
about 3\% of the CCD images have a limiting magnitude
above that needed to detect clusters out to $z_{est} = 1.0$ and
about 20\% have insufficient depth to detect clusters in the
range $1 < z_{est} \le 1.2$.

\subsection{The Space Density of Clusters}
\label{spaceden}

\begin{figure*}[t]
\leavevmode
\epsfxsize=6.5inch
\epsfbox{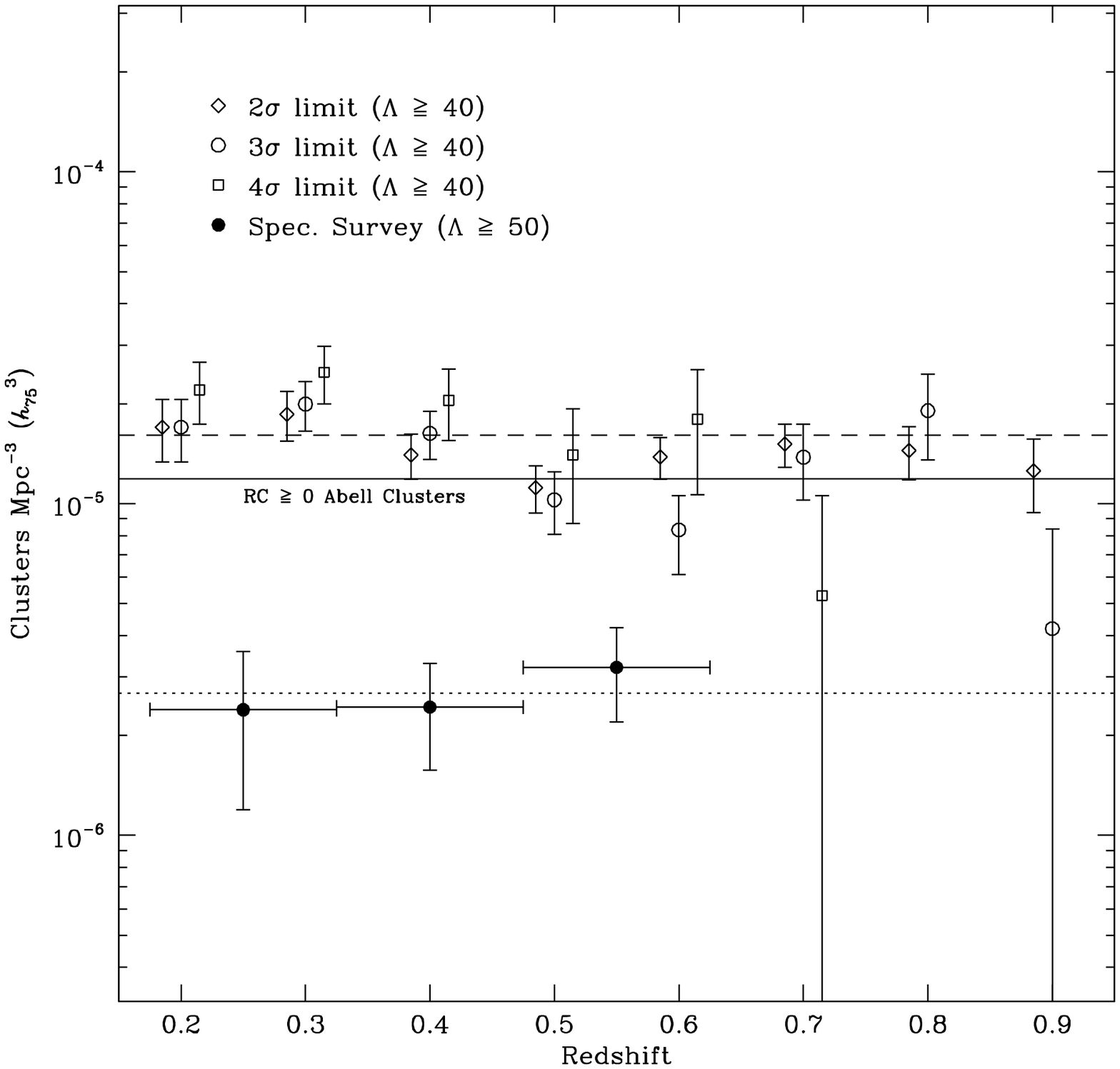}
\caption{
The space density of clusters as a function of redshift.
The open diamonds, circles, and squares show the space density as
a function of estimated redshift for all cluster
candidates with $\Lambda_{CL} \ge 40$ and with detection
thresholds of 2, 3, and 4$\sigma$, respectively. These data have
been divided by the appropriate selection functions and corrected
for the expected false positive rates. The data points for the
2 and 4 sigma results are intentionally offset (by $\pm0.015$ in $z$) for
clarity. The solid circles show the space density averaged
into 3 redshift bins for the 22
$\Lambda_{CL} \ge 50$ clusters with 4 or more concordant spectroscopic
redshifts. The long-dashed line is the mean space density for the
$\Lambda_{CL} \ge 40$ subsample, the short-dashed line is the mean
space density for the spectroscopic sample. The solid line is
the mean space density of RC $\ge 0$ Abell clusters.}
\label{figspaceden}
\end{figure*}

We use the estimated redshifts in the full catalog to derive the 
space density of clusters as a function of redshift. At each redshift,
the space density is 
\begin{equation}
\rho(z) = \Bigl(\sum\limits_{k=1}^{N_{CL}(z)} 1/P_k(z) - N_{spur}(z)\Bigr) \Biggl/ \int\limits_{z-\Delta z/2}^{z+\Delta z/2} \Omega(z) dV
\end{equation}
where $P(z)$ is the probability of detecting a cluster with a given
$\Lambda_{CL}$ and detection threshold at redshift $z$, $N_{CL}(z)$ is the
number of clusters found with estimated redshift between 
$z - \Delta z/2$ and $z + \Delta z/2$, $N_{spur}(z)$ is the mean number of false positive
detections expected in this same redshift interval, $\Omega(z)$ is the effective
area subtended by the survey at redshift $z$, and $dV$ is the
unit comoving volume element. The width of the redshift
bins, $\Delta z$, is 0.1. 
Figure~\ref{figspaceden} shows the space density of $\Lambda_{CL} \ge 40$
clusters as a function of redshift and detection significance. The minimal 
dependence of the cluster space density on detection significance is a tribute to the
accuracy of the selection functions -- weighting the cluster counts by
the inverse of the detection probability effectively removes the
artificial decline in space density introduced by the decreasing
significance of the cluster detections with increasing redshift.
The mean space density of $\Lambda_{CL} \ge 40$ clusters, in the range
$0.2 \le z_{est} < 1$, is $(1.61 \pm 0.24) \times 10^{-5}h_{75}^{3}$ Mpc$^{-3}$,
in excellent agreement with the best-fit space density obtained in \S\ref{surfden}. 
The mean $V/V_{max}$ value \citep{schmidt} for the $\Lambda_{CL} \ge 40, \sigma \ge 3.0$
cluster sample is 0.54, suggesting that the spatial distribution of
clusters this sample is close to uniform and, thus, the  
space densities derived from it are representative of the full redshift range.
The observed comoving cluster space density is consistent with little or no
evolution over the redshift range $0.2 < z < 0.8$. Attempting to fit an exponentially
decaying space density to the data allows us to place an upper limit to any density
evolution. A drop in space density of greater than a factor of 3 over the above
redshift range is ruled out at greater than the 99.9\% confidence level for the  
sample of clusters with $\Lambda_{CL} \ge 40,\ \sigma \ge 3$.

Figure~\ref{figspaceden} also shows the space density of $\Lambda_{CL} \ge 50$ clusters
that have 4 or more concordant spectroscopic redshifts. The
mean space density for these 22 clusters is 
$(2.68 \pm 1.05) \times 10^{-6}h_{75}^{3}$ Mpc$^{-3}$.
If we include those clusters with at least 3 concordant redshifts,
then the mean space density becomes $(3.38 \pm 1.17) \times 10^{-6}h_{75}^{3}$ Mpc$^{-3}$.
These densities differ by a factor of $\sim 3$ from the result obtained
using the full $\Lambda_{CL} \ge 50$ sample with estimated redshifts: 
$(8.80 \pm 0.13) \times 10^{-6}h_{75}^{3}$ Mpc$^{-3}$. About 80\% of
the difference is due to the significant volume underestimation that results from 
deriving the space density for a sample spanning a very narrow range of
estimated redshifts. If we recompute the space density of the spectroscopically
observed clusters using their estimated redshifts instead, we derive density
that is 2.4 times higher than the space density derived using the spectroscopic
redshifts. This systematic error is a direct consequence of the moderate uncertainties
in the redshift estimates and was discussed originally in \citet{p96}.
The 27 clusters with existing redshifts in the spectroscopic followup survey  
all have estimated redshifts of either 0.3 or 0.4, yet their spectroscopic
redshifts span the range $0.225 \le z \le 0.611$. The remaining difference
($\sim20$\%) is probably due to small uncertainties in our false positive rate
corrections and incompleteness in our spectroscopic survey (the S/N ratios of the
spectra in 3 of the 36 cluster candidates in the spectroscopic survey 
were too low to derive redshifts). 

In general, the effect of
uncertainties in the estimated redshift and uncertainties in $\Lambda_{CL}$
conspire to inflate the derived abundance of clusters by some amount that is
dependent on the amplitude of these uncertainties. This is largely due to
an asymmetry in the number of clusters that are scattered into a given
estimated redshift bin versus the number scattered out. However,
the above factor of $\sim 2.4$ volume underestimate is not applicable 
across the full estimated redshift range $0.2 \le z_{est} \le 1.2$. 
Monte Carlo simulations we have performed indicate that the mean 
space density derived from
a sample that covers an estimated redshift range of at least 5 times the rms
scatter in the $(z_{obs} - z_{est})$ values should 
be no more than $\sim15$\% larger than the true space density. 
We have corrected all our space density estimates downward accordingly to account 
for this effect. The spectroscopic sample only 
spans an estimated redshift range that is 
comparable to this rms scatter (see \S\ref{zest}) 
and, thus, the significant volume underestimate resulting
from use of the estimated redshifts is fully expected.
The corrected mean space density derived from the full
survey ($0.2 \le z_{est} \le 1.2$) should, therefore, be relatively free from
systematic error unless the rms scatter in the estimated redshifts is significantly
larger than 0.2 -- and this is not supported by existing spectroscopic data
(see \S\ref{zest}). 
 
\subsection{The Lambda Function and Cluster Mass Estimation}
\label{lamfunc}

The dependence of the space density of clusters on $\Lambda_{CL}$ is
equally as important as its redshift dependence. In the absence of an accurate
mass estimate for each cluster, the $\Lambda_{CL}$ function can provide a proxy
for the mass function if the mass-to-light ratio on cluster scales is relatively
invariant or if the $\Lambda_{CL}$ value is based on a rest-wavelength that traces
the integrated stellar mass in the cluster galaxies. In the case of the Deeprange
survey, the latter is certainly not the case for redshifts higher than $z \sim 0.4$. 
Nonetheless, the relative
abundance of clusters as a function of $\Lambda_{CL}$ can be used to provide
a crude constraint on the typical mass of the ensemble of clusters in the survey,
as discussed at the end of this section.

\begin{figure*}[t]
\leavevmode
\epsfxsize=6.5inch
\epsfbox{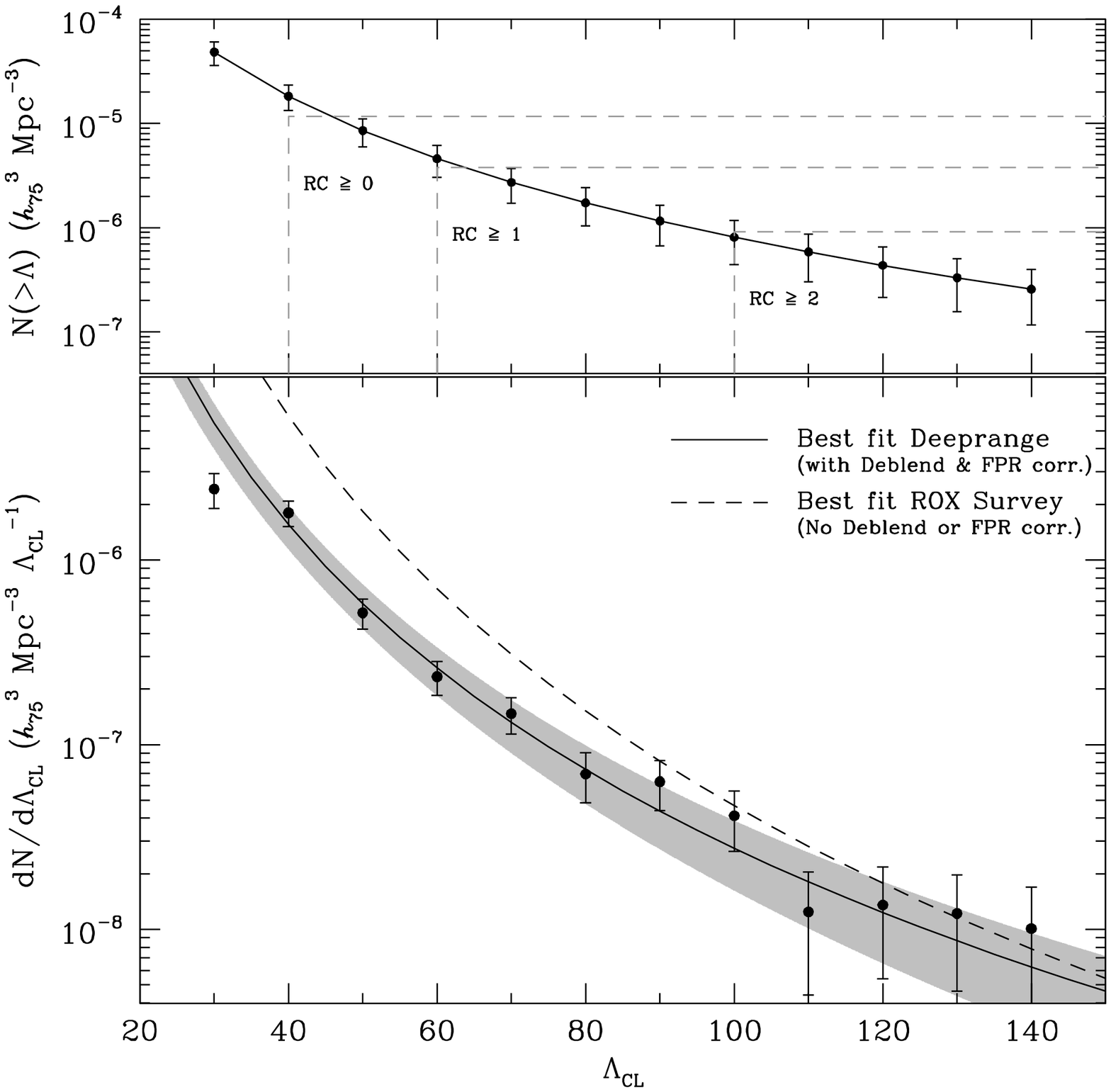}
\caption{{\bf Lower panel:} The differential
space density of clusters as a function of $\Lambda_{CL}$
for the subset of cluster candidates detected with at least a
3$\sigma$ significance. The cluster counts are weighted by the appropriate
selection function and corrected for the expected false positive rate.
The grey shaded region shows the 1$\sigma$ confidence
limits on the best-fit power-law relation. The best-fit from the ROXS
data are shown for comparison. The difference between our result and
the ROXS data is understood and discussed in \S\ref{lamfunc}.
{\bf Upper Panel:} The integrated space density of clusters derived from
the best-fit power-law to the Deeprange data in the lower panel. The
dashed lines correspond to the observed Abell
cluster space densities as a function of richness.}
\label{figlamfunc}
\end{figure*}

We use the same procedures for computing the differential $\Lambda_{CL}$ function 
as in \citet{rox01} for the ROSAT Optical X-Ray Survey (ROXS).
In our case, we calculate the $\Lambda_{CL}$ function for the subset 
of 217 cluster candidates with detection significance $>3\sigma$
and $\Lambda_{CL} \ge 20$.
The data and best-fit are shown in the lower panel of Figure~\ref{figlamfunc}.
The data points shown are the binned values in
14 equal sized $\Delta\Lambda_{CL}=10$ intervals such
that $dN/d\Lambda_{CL} = \sum_{i}1/V_{max,i}$ where 
$V_{max}$ is the maximum volume at which a cluster could have been
detected in our survey. The $V_{max}$ for a given cluster is computed by 
integrating the cosmological comoving volume element $dV$ multiplied
by the appropriate detection probability function. For this purpose, 
the $\Lambda_{CL}-$dependent selection functions (see Figure~\ref{figselfunc})
were interpolated to span the range of $\Lambda_{CL}$ values in the above subsample.

In order to characterize the general shape and normalization of
the $\Lambda_{CL}$ function, the unbinned $\Lambda_{CL}$ and estimated redshift 
data were fit to a simple power law of the form 
$dN/d\Lambda_{CL} = N_0 (\Lambda_{CL}/40)^{-\alpha}$, where the normalization
$N_0$ and the slope $\alpha$ were the sole parameters of the fit.
We compute a likelihood function $S(\alpha,N_0)$ \citep{matz} defined by
\begin{equation} 
S(\alpha,N_0) = 2 \int N_{0} (\Lambda_{CL}/40)^{-\alpha} f(\Lambda_{CL},z) 
d\Lambda_{CL} dV -2 \log \cal{N} 
\end{equation}
where $\cal{N}$ is a sum of all $k$ cluster candidates (appropriately weighted
to account for the expected false positive rate)
such that $\cal{N}$ $ = \sum_{k} N_{0} (\Lambda_{CL,k}/40)^{-\alpha}$ 
and $f(\Lambda_{CL},z)$ is the normalized selection function
(Figure~\ref{figselfunc}) assuming elliptical-like passive evolution
and a cluster profile slope of $\gamma=-1.4$.
The best-fit function for the 217 clusters with detection significance of at least
$3\sigma$ and $\Lambda_{CL} \ge 20$ is
\begin{equation}
dN/d\Lambda_{CL} = (1.55 \pm 0.40) \times 10^{-6} (\Lambda_{CL}/40)^{-4.40 \pm 0.30}
\end{equation}
The units of $dN$ are $h_{75}^{3}$ Mpc$^{-3}$, $\Lambda_{CL}$ is dimensionless.
The uncertainties in the normalization and slope are statistical only
and do not include systematic uncertainties associated with 
false positive rates, volume/redshift estimates, and $\Lambda_{CL}$.
We further caution that the best-fit is not intended for extrapolation beyond the
redshift or $\Lambda_{CL}$ ranges spanned by the data in Figure~\ref{figlamfunc}. 

Formally, the ROXS data \citep{rox01} resulted in a steeper 
$\Lambda_{CL}$ function 
($\alpha = 4.8-5.8$, $1\sigma$) with higher normalization at $\Lambda_{CL}=40$
(by factor of $\sim2$). The difference is largely for cluster candidates
in the $\Lambda_{CL} < 60$ range and is most likely due to differences in the
processing of the ROXS data -- \citet{rox01} do not correct their
counts for false positive detections nor do they ``deblend"
their cluster detections, which results in a higher number of low-$\Lambda_{CL}$
systems in their catalog. 
Our $\Lambda_{CL}$ function fit is not inconsistent with the ROXS data, however. The
slopes of the two Lambda functions differ by a formal uncertainty of $2\sigma$.
Since most of the systematic uncertainties in the data 
induce dispersion, the discrepancy between the two $\Lambda_{CL}$ function fits is,
in fact, less significant than $2\sigma$.

The cumulative cluster space densities, $N(\ge \Lambda_{CL})$, derived 
by integrating equation 9 agree with those derived in \S\ref{surfden} and 
\S\ref{spaceden} to within $\pm 25$\%, well within 
the statistical uncertainties of the fit.
For $\Lambda_{CL} \ge 40$, the integration yields a space density
of $(1.83 \pm 0.50) \times 10^{-5}h_{75}^{3}$ Mpc$^{-3}$, about
1.6 times the density of Abell RC $\ge 0$ clusters.
Table 7 lists the integrated space densities as a function of $\Lambda_{CL}$
and these data are plotted in the upper panel of Figure~\ref{figlamfunc}.
Our $\Lambda_{CL} \ge 40$ space density is also consistent with
that seen in recent x-ray selected surveys for local clusters.
For example, integrating the \citet{bcs97} bolometric x-ray luminosity function
one obtains a space density of $1.57 \times 10^{-5}h_{75}^{3}$ Mpc$^{-3}$
for clusters with $L_x \ge 2.25 \times 10^{43}h_{75}^{-2}$ erg s$^{-1}$,
the typical x-ray luminosity for a $\Lambda_{CL} = 50$ cluster \citep{rox01}.
Our calibration of the $\Lambda_{CL} - N_A$ relation predicts that $\Lambda_{CL} \ge 40$
corresponds to RC $\ge 0$ and the uncertainties in our false positive
rate correction are no larger than 20\%. Therefore, the most probable explanation
for the lower space density of Abell clusters 
is likely to be attributed to incompleteness in low richness
end of the Abell catalog. The space density of $\Lambda_{CL} \ge 60$ clusters
in our survey agrees with the local space density RC $\ge 1$ Abell clusters
to within 20\%. The cumulative Abell cluster space densities for
richness class RC $\ge 0$, RC $\ge 1$, and RC $\ge 2$
are displayed in the upper panel of Figure~\ref{figlamfunc}
as dashed lines.

We can use the $\Lambda_{CL}$ function to obtain approximate estimates of 
the typical cluster mass as a function $\Lambda_{CL}$. The relation
between $\Lambda_{CL}$ and mass is expected to be noisy \citep{rox01, rox02}.
As such, one cannot reliably take a $\Lambda_{CL}$ value for an individual 
cluster and expect to derive an accurate mass. One can, however, attempt to
provide an estimate of the typical mass for a large ensemble of clusters
in at least two independent ways. In the first method, we compute the average
$\Lambda_{CL}$ value for a subset of clusters, derive the corresponding
average intrinsic luminosity (recall that $\Lambda_{CL}$ corresponds to the 
effective number of $L^{*}$ galaxies within the cutoff radius used in the matched filter),
correct this luminosity for the finite radius used, and multiply the
corrected luminosity by an assumed M/L ratio to obtain a mean mass estimate
corresponding to the mean $\Lambda_{CL}$.
The second method involves deriving a predicted mass function using
Press-Schechter formalism for a given cosmological model and finding
the correspondence between the densities in the model with the observed
cluster space densities (\eg Table 7).

We use the first method to derive mass estimates for clusters with
$\Lambda_{CL} \ge$ 40, 60, and 100, respectively, which correspond
approximately to the Abell richness class limits $\ge 0$, $\ge 1$, and $\ge 2$.
The mean $\Lambda_{CL}$ values for these three samples are 57, 85, and 141.
We convert these mean $\Lambda_{CL}$ values to absolute luminosities using our 
adopted $M^{*}_I$ value of $-21.90$ ($2.1 \times 10^{10} h_{75}^{-2} L_{\odot}$)
and multiply by a factor of 1.05 to correct for the finite radius\footnote{For
a de-projected density profile with a core radius of 133 kpc, 
the ratio between the integrated mass at infinite radius
and that at the matched filter cutoff radius is 1.02 and 1.07 for
profile slopes of -2.4 and -2.0, respectively.} of the matched filter. 
The finite radius correction factor is, however, considerably smaller 
than the uncertainties in the M/L ratio.
If $\Omega_m = 0.2$, then the typical M/L ratio on cluster scales in the
$I-$band should be $\sim200$, based on the luminosity function
parameters presented in \citet{blan}. If the M/L is independent
of mass, then we find that the typical cluster masses for 
$\Lambda_{CL} \ge$ 40, 60, and 100 are
$2.5 \times 10^{14}$M$_\odot$,
$3.7 \times 10^{14}$M$_\odot$, and
$6.2 \times 10^{14}$M$_\odot$, respectively. If the M/L increases
with cluster mass as $M/L \propto M^{\sim0.5}$ 
\citep{hr00, rox01} then the mass estimates
above would be correspondingly higher by $\sim 60 - 160$\%.

In the second method, a Press-Schechter mass-function is derived from
the approximation developed by \citet{pen} for a flat-universe with
$\Omega_m = 0.3$, $\Omega_b h^2 = 0.0205$ \citep{om01},
$h = 0.75$, and $\sigma_8 = 1.0$, which is consistent results from 
cluster abundance constraints (\eg \citealt{mos01}).
In this model, $N(>M) = 10^{-6}h_{75}^{3}$ Mpc$^{-3}$
when $M \approx 3 \times 10^{14}$M$_\odot$. This would correspond
to the observed cluster density when $\Lambda_{CL} \gsim 90$. This
mass scale is in good agreement with that derived in the 
first approach. While quite approximate, both approaches suggest that
the average mass of a $\Lambda_{CL} \ge 40$ cluster is in the
few $\times 10^{14}$M$_\odot$ range, albeit the scatter about this mean
value is expected to be substantial. 

\subsection{Superclusters at Intermediate Redshifts}
\label{sclus}

We have identified at least 2 potential superclusters from our spectroscopic 
survey data. The clusters CL416, CL440, and CL441 are linked when a percolation
length of 10$\mpc$ (${\delta\rho \over \rho} \approx 100$) is used. The mean redshift
of this supercluster is 0.482. The clusters CL426 and CL433 are also linked with this
percolation length and lie at a mean redshift of 0.226. At a slightly larger 
percolation of 13$\mpc$ (${\delta\rho \over \rho} \approx 50$), the clusters
CL415 and CL439 are linked. The mean redshift of this system is 0.357.
No additional supercluster candidates are found for $10 < {\delta\rho \over \rho} < 50$.
This supercluster frequency is comparable to that in the
distribution of local RC $\ge 0$ Abell clusters for similar overdensities --
$\sim10 - 20$\% of Abell clusters are linked into groups of 3 or more
when $10 < {\delta\rho \over \rho} < 20$ \citep{p92, lp94}.
The frequency of superclusters at a given
overdensity is correlated with the cluster-cluster correlation length. As our 
spectroscopic survey is largely complete for $\Lambda_{CL} \ge 50$ clusters with 
$0.3 \le z_{est} < 0.5$, the general agreement between our supercluster frequency 
and that in the Abell catalog suggests that the correlation length for the 
$\Lambda_{CL} \ge 50$ systems at $z \sim 0.4\pm0.1$ is similar to the local value
for RC $\ge 0$ clusters. If the cluster distribution at intermediate redshifts
were completely random, the probability of detecting a supercluster with
at least 3 members and with ${\delta\rho \over \rho} > 10$ would be less than 0.005.
Clearly, this is a crude qualitative constraint on the evolution of 
clustering but it is consistent with the findings of \citet{gonz}.
who find only modest evolution in the cluster-cluster correlation length
by $z \sim 0.45$ from the Las Campanas Distant Cluster Survey.

\section{Conclusions}
\label{conc}

We have completed an automated search for distant clusters in a contiguous
16 square degree patch of sky using moderately deep $I-$band images as our
source data. The complete cluster catalog and redshifts for those clusters with
spectroscopic data are provided. The key results derived in this paper are
summarized below. The images from the Deeprange survey are publicly available
from the NOAO science archive (http://archive.noao.edu/nsa/).

({\it i}) Various approaches to estimating the
space density of clusters in our deep survey yield 
values in the range $(1.57 - 1.83) \times 10^{-5}h_{75}^{3}$ Mpc$^{-3}$
for systems with $\Lambda_{CL} \ge 40$. The uncertainties in these
estimates range from 10\% to 27\%, depending on the technique used.
This space density range is about a factor of $1.3 - 1.6$ times 
larger than that for comparably rich clusters
(RC $\ge 0$) in the Abell catalog.
The offset between our results and those for the Abell catalog is due predominantly
to differences at the low richness end of the cluster distribution.
The ratio of Deeprange-to-Abell cluster space density
drops to 1.2 for $\Lambda_{CL} \ge 60$ (RC $\ge 1$) and 
is consistent with unity for $\Lambda_{CL} \ge 100$ (RC $> 2$). 
The discrepancy between the Abell catalog and our current survey is substantially
less than that reported for the Palomar Distant Cluster Survey \citep{p96}.
We believe this is mostly do to the use of better values for the local space
density of clusters and a more reliable calibration of the
$\Lambda_{CL} - N_A$ relation in the current work. The cluster abundances
found here are also consistent with the density of X-ray selected clusters
in wide area surveys (\eg \citealt{bcs97}). \citet{br00} apply matched
filter cluster detection to the Edinburgh/Durham Cluster Catalog II (EDCCII) and find
a significantly higher local space density of clusters than is found in the
Abell catalog: $2.4^{+2.7}_{-1.0} \times 10^{-5}h_{75}^{3}$ Mpc$^{-3}$ 
for $\Lambda_{CL} \ge 40$ systems. However, their richness calibration 
associates $\Lambda_{CL} \ge 40$ with
Abell RC $\ge 1$, in part due to their use of the now superseded cluster space
densities given in \citet{b79}. Our revised calibration would associate such systems 
with poorer $RC \ge 0$ clusters. If we revise their calibration to 
match ours then we find
good agreement between their local space density and that seen at $z > 0.2$.

({\it ii}) The comoving space density of clusters with $\Lambda_{CL} \ge 40$ is
relatively constant out to $z=0.8$. While the data do allow a gradual decline
in cluster abundance with increasing redshift, a drop in density by a factor
greater than 3 over the range $0.2 < z < 0.8$ is ruled out at $>$99.9\% 
confidence level. \citet{p96} and \citet{bcs98} have previously
reported results suggesting non-evolving cluster space densities out to
$z = 0.6$ and $z = 0.3$, respectively. The mean $V/V_{max}$ value for the
$\Lambda_{CL} \ge 40$ sample is 0.54, in excellent agreement with the
expected value for a uniform, non-evolving distribution. Coupled with the
good agreement between our results and those from local x-ray surveys (\eg
\citealt{bcs97}) and a revised estimate of the density found in the EDCCII catalog 
\citep{br00}, a scenario where 
the space density of clusters has evolved very little, if at all, over the 
past half a Hubble time appears to be likely. 
This result favors CDM models with low values of $\Omega_m$. A thorough consistency
check between these data and model predictions will be the scope of a future paper.
We note, however, that such a comparison is complicated by the fact that what we are
measuring here is the number density of clusters as a function of optical luminosity
and redshift whereas what models most often predict is the evolution of the 
cluster mass function. A proper analysis, therefore, either requires an accurate
estimate of the mass distribution of these clusters or a prediction of the evolution of
the space density of clusters in a rest-frame optical passband. Furthermore,
\citet{voit} demonstrates that the evolution of the mass function of clusters
and the evolution of their ICM temperature (which correlates with x-ray luminosity)
differ even for a given cosmological model. Specifically, the temperature function
evolution tends to be weaker than the mass function evolution. If similar trends
are carried into the optical bandpass, albeit driven by different astrophysical
processes, the weak (or non-existent) variation
of cluster space density with redshift observed here may not imply a similarly
weak evolution in the cluster mass function. What is certain, however, is that in an
$\Omega_m = 1$ universe the temperature function evolves dramatically -- the normalization
increases by at least 3 orders of magnitude between $z = 1$ and $z = 0$. Even with
a noisy correlation between optical and x-ray luminosity, such substantial
evolution is clearly ruled out by the Deeprange survey. 
 
({\it iii}) An empirical calibration of the matched filter cluster luminosity, 
$\Lambda_{CL}$, and the Abell richness parameter yields $\Lambda_{CL} = 1.24 N_A$. 
This calibration agrees well with the relationship predicted from extensive simulations.

({\it iv}) The false positive rate using the matched filter detection method on
these data is, on average, between 20 -- 30\%. This assessment is based on a
spectroscopic follow-up survey of $\sim40$ cluster candidates in the
range $0.2 < z < 0.6$ and on simulations explicitly designed to estimate
the spurious detection rate. The false positive rate is a function of 
effective cluster size, detection significance, and cluster richness. 
Cluster samples that are largely free of spurious detections can be
constructed by appropriately filtering on these parameters. 

({\it v}) The $\Lambda_{CL}$ function is consistent with a power law of the form 
$dN/d\Lambda_{CL} = (1.55 \pm 0.40) \times 10^{-6} (\Lambda_{CL}/40)^{-4.40 \pm 0.30}$, 
where $dN$ is in units of $h_{75}^{3}$ Mpc$^{-3}$.
Two separate methods to provide an approximate calibration of the
mass of the clusters in this survey find that 
the average mass of a $\Lambda_{CL} \ge 40$ cluster is in the
few $\times 10^{14}$M$_\odot$ range. However, the mass-$\Lambda_{CL}$ relation
is too noisy to provide accurate estimates of individual cluster masses.

({\it vi}) Between 10 -- 20\% of the spectroscopically confirmed clusters are
linked into superclusters at high overdensities (${\delta\rho \over \rho} > 10$).
This percentage is comparable with the value derived from local cluster surveys
and suggests that the cluster -- cluster correlation function does not evolve
dramatically between $z \sim 0.5$ and the current epoch.

\acknowledgments

We thank the NOAO and U.S. Gemini TACs for their support of this survey.
We thank Rick White for his assistance in enhancing the astrometric 
accuracy of our coordinates via cross-correlation with the FIRST survey. 
We thank the anonymous referee for helpful comments, which improved the
clarity of the paper.
We also thank Lori Lubin for valuable comments on the initial draft of 
this paper.

\clearpage
\begin{deluxetable}{ll}
\tablewidth{0pt}
\tablenum{1}
\tablecaption{Cluster Detection Parameters}
\tablehead{
Parameter &
Value}
\startdata
Hubble constant      & 75 km s$^{-1}$ Mpc$^{-1}$ \\
$\Omega_m$           & 0.2 ($\Lambda = 0$)       \\
Core Radius          & 133 kpc                   \\
Cutoff Radius        & 1.33 Mpc                  \\
LF slope             & -1.10                     \\
LF M$^*$ ($z=0$)     & -21.90                    \\
Detection Threshold  & 2$\sigma$                 \\
Min. Filter Redshift & 0.20                      \\
Max. Filter Redshift & 1.20                      \\
\enddata
\end{deluxetable}

\clearpage
\hoffset -0.5in
\begin{deluxetable}{c|rr|r|r|rrrr|rr|rr|c}
\tablewidth{0pt}
\tabletypesize{\scriptsize}
\tablenum{2}
\tablecaption{Deeprange Cluster Catalog}
\tablehead{
 & & & & & & & & & & Rad & & & \\
\multicolumn{1}{c|}{ID} &
\multicolumn{2}{c|}{R.A. (J2000) Dec.} &
\multicolumn{1}{c|}{$z_{\rm est}$} &
\multicolumn{1}{c|}{$\Lambda_{\rm CL}$} &
\multicolumn{1}{c}{N$_{A,0.25}$} &
\multicolumn{1}{c}{N$_{e,0.25}$} &
\multicolumn{1}{c}{N$_{A,0.5}$} &
\multicolumn{1}{c|}{N$_{e,0.5}$} &
\multicolumn{1}{c}{$\sigma$} &
\multicolumn{1}{c|}{$('')$} &
\multicolumn{1}{c}{m$_3$} &
\multicolumn{1}{c|}{m$_{eff}$} &
\multicolumn{1}{c}{QA} }
\startdata
001&10 08 40.77&$+$54 32 49.7&0.60& 43.0& 10.0&  7.7& 22.0& 25.6& 2.1&  14&20.8&21.3& \nodata \\
002&10 06 30.68&$+$54 32 29.1&0.20& 23.6&  8.4&  7.4&  6.5& -0.3& 3.0&  61&17.8&18.4& \nodata \\
003&10 17 44.67&$+$54 32 37.5&0.30& 28.6& 10.7& 13.2& -7.3&-10.2& 2.6&  45&19.0&19.5& \nodata \\
004&10 22 52.82&$+$54 32 29.4&0.80& 65.5& 19.6& 14.3& 27.5& 19.0& 2.3&  30&21.0&22.1&2\\
005&10 10 58.60&$+$54 33 30.9&0.30& 53.0& 11.3& 15.3& 22.0& 59.1& 4.8& 252&18.0&19.3&1\\
006&10 12 49.53&$+$54 35 03.8&0.30& 40.3& 11.5&  4.3&  2.0& -8.9& 3.7& 180&18.8&19.2& \nodata \\
007&10 17 00.50&$+$54 30 46.2&1.10& 91.5& 12.2&  8.4& 16.3&  5.4& 2.1&  18&22.0&23.0&1\\
008&10 25 29.15&$+$54 34 55.6&0.20& 36.7&  4.4&  5.4& -2.3& -6.5& 4.6& 195&17.3&17.8& \nodata \\
\enddata
\tablecomments{   CL309 = Abell ~~985, $N_A = 61$, $z = 0.1347$ (redshift from NED)\\
\phantom{XXXXXXX} CL414 = Abell 1004,  $N_A = 76$, $z = 0.1414$ (redshift from NED)\\
\phantom{XXXXXXX} CL423 = Abell ~~915, $N_A = 54$, $z = 0.1484$ (redshift from NED)}
\end{deluxetable}

\clearpage
\begin{deluxetable}{cccccc}
\tablewidth{0pt}
\tablenum{3}
\tablecaption{Cluster Sample Size vs. Minimum Selection Criteria}
\tablehead{
\multicolumn{1}{c}{~} &            
\multicolumn{1}{c}{~} &
\multicolumn{1}{c}{Min. Effective} &
\multicolumn{1}{c}{Number}               &
\multicolumn{1}{c}{No. with} &
\multicolumn{1}{c}{No. with} \\
\multicolumn{1}{c}{Min. $\sigma$} &
\multicolumn{1}{c}{Min. $\Lambda_{CL}$} &
\multicolumn{1}{c}{Radius (arcsec)} &
\multicolumn{1}{c}{of Clusters} &
\multicolumn{1}{c}{$z_{est} \ge 0.7$} &
\multicolumn{1}{c}{$z_{est} \ge 1.0$} }
\startdata
2.0 & 10   & 5   & 444 & 154 & 63 \\
2.5 & 10   & 5   & 341 &  91 & 33 \\
3.0 & 10   & 5   & 217 &  38 & 10 \\
3.5 & 10   & 5   & 146 &  16 &  2 \\
4.0 & 10   & 5   &  95 &   2 &  1 \\
    &      &     &     &     &    \\
2.0 & 10   & 5   & 444 & 154 & 63 \\
2.0 & 20   & 5   & 440 & 154 & 63 \\
2.0 & 40   & 5   & 338 & 154 & 63 \\
2.0 & 60   & 5   & 189 & 134 & 63 \\
2.0 & 80   & 5   & 106 &  92 & 61 \\
    &      &     &     &     &    \\
2.0 & 10   &  5  & 444 & 154 & 63 \\
2.0 & 10   & 10  & 441 & 151 & 60 \\
2.0 & 10   & 25  & 402 & 122 & 48 \\
2.0 & 10   & 50  & 329 &  89 & 32 \\
2.0 & 10   & 75  & 267 &  70 & 24 \\
    &      &     &     &     &    \\
3.0 & 60   & 60  &  93 &  38 & 10 \\
\enddata
\end{deluxetable}

\clearpage
\begin{deluxetable}{l|rr|r|cr|l}
\tablewidth{0pt}
\tabletypesize{\scriptsize}
\tablenum{4}
\tablecaption{Galaxy Redshifts}
\tablehead{
\multicolumn{1}{c|}{~} & 
\multicolumn{2}{c|}{~} & 
\multicolumn{1}{c|}{~} & 
\multicolumn{1}{c}{~} & 
\multicolumn{1}{c|}{$z$ Error} & 
\multicolumn{1}{c}{~} \\
\multicolumn{1}{c|}{ID} & 
\multicolumn{2}{c|}{R.A. (J2000) Dec.} & 
\multicolumn{1}{c|}{I mag} & 
\multicolumn{1}{c}{$z$} & 
\multicolumn{1}{c|}{(km/s)} & 
\multicolumn{1}{c}{Comments} }
\startdata
CL005 & & & & & & \\
1 & 10 10 45.21&+54 34 45.2&19.73&\nodata &\nodata & Low S/N \\
2 & 10 10 51.25&+54 35 09.6&20.23&0.3987 &153 & \\
3* & 10 10 50.83&+54 34 24.8&19.03&0.3629 &138 & \\
4 & 10 10 46.94&+54 33 12.9&19.78&\nodata &\nodata & Low S/N\\
5* & 10 11 01.52&+54 35 19.0&20.45&0.3654 &174 & \\
6 & 10 10 57.68&+54 34 05.7&18.34&0.2543 & 140& Emission\\
7* & 10 11 01.00&+54 33 33.5&19.10&0.3650 &120 & \\
\enddata
\tablecomments{Galaxies with an asterisk after their ID number are
used in the computation of the mean cluster redshift. See Table 5
for results.}
\end{deluxetable}

\clearpage
\begin{deluxetable}{llrrr|llrrr}
\tablewidth{0pt}
\tablenum{5}
\tablecaption{Cluster Redshifts}
\tablehead{
\multicolumn{1}{c}{Cluster} & 
\multicolumn{1}{c}{$\overline z$} & 
\multicolumn{1}{c}{$N_z$} & 
\multicolumn{1}{c}{$N_{spec}$} & 
\multicolumn{1}{c|}{$N_{aper}$} &
\multicolumn{1}{c}{Cluster} & 
\multicolumn{1}{c}{$\overline z$} & 
\multicolumn{1}{c}{$N_z$} & 
\multicolumn{1}{c}{$N_{spec}$} & 
\multicolumn{1}{c}{$N_{aper}$} }
\startdata
CL005  & 0.3644   & 3       &  6  & 11 &
CL310  & 0.5005 $\dagger$& 3&  8  & 10 \\
CL009  & 0.5543   & 3       &  7  & 13 &
CL335  & 0.6111   & 3       &  7  & 12 \\
CL011  & 0.3026   & 3       &  7  &  8 &
CL338  & 0.2544   & 4       &  4  & 10 \\
CL026  & 0.5795 ? & 2       &  4  &  9 &
CL350  & 0.3925 $\dagger$& 3&  6  &  8 \\
CL052  & 0.5444   & 5       &  7  & 10 &
CL378  & 0.5147   & 4       &  5  &  8 \\
CL062  & 0.4615 ? & 2       &  5  & 11 &
CL384  & 0.5021   & 5       &  8  &  9 \\
CL063  & 0.3292   & 6       & 10  & 11 &
CL385  & 0.2526 ? & 2       &  5  &  9 \\
CL065  & 0.5002   & 5       &  9  & 11 &
CL397  & 0.4914   & 4       &  5  & 12 \\
\enddata
\tablecomments{$N_z$ is the number of galaxies used to compute the 
cluster redshift, $N_{spec}$ is the number of galaxy redshifts measured,
$N_{aper}$ is the number of spectra extracted. See Table 4 for the specific
galaxies used in the $\overline z$ computation. Cluster redshifts marked with
the $\dagger$ symbol have a maximum spread in the individual galaxy redshifts 
that exceeds the nominal 3000 km s$^{-1}$ limit. Redshifts with question marks
are based on fewer than 3 galaxies and may not be very reliable.}
\end{deluxetable} 

\clearpage
\begin{deluxetable}{crc}
\tablewidth{0pt}
\tablenum{6}
\tablecaption{$\Lambda_{CL} \ge 40$ Cluster Surface Density, Corrected for Selection Function}
\tablehead{
\multicolumn{1}{c}{~}         &
\multicolumn{1}{c}{Clusters deg$^{-2}$} &
\multicolumn{1}{c}{Effective Area} \\
\multicolumn{1}{c}{$z_{est}$} &
\multicolumn{1}{c}{with $z \le z_{est}$} &
\multicolumn{1}{c}{of Survey (deg$^2$)}}
\startdata
0.20 & $ 1.19 \pm 0.28$ & 14.762 \\
0.30 & $ 3.02 \pm 0.46$ & 14.762 \\
0.40 & $ 5.18 \pm 0.64$ & 14.762 \\
0.50 & $ 7.52 \pm 0.85$ & 14.762 \\
0.60 & $10.69 \pm 1.12$ & 14.586 \\
0.70 & $14.70 \pm 1.47$ & 14.586 \\
0.80 & $19.31 \pm 1.88$ & 14.528 \\
0.90 & $24.46 \pm 2.37$ & 14.528 \\
1.00 & $31.02 \pm 2.97$ & 14.293 \\
1.10 & $39.81 \pm 3.74$ & 13.005 \\
1.20 & $47.97 \pm 4.50$ & 11.071 \\
\enddata
\end{deluxetable}

\clearpage
\begin{deluxetable}{rc|c}
\tablewidth{0pt}
\tablenum{7}
\tablecaption{Cumulative Cluster Space Densities}
\tablehead{
\multicolumn{1}{c}{~} &
\multicolumn{1}{c}{$N(\ge \Lambda_{CL}$)} &
\multicolumn{1}{|c}{Corresponding Abell}\\
\multicolumn{1}{c}{$\Lambda_{CL}$} &
\multicolumn{1}{c}{$h_{\rm 75}^{3}$ Mpc$^{-3}$} &
\multicolumn{1}{|c}{Cluster Density}}
\startdata
  30 & $(4.85 \pm 1.25) \times 10^{-5}$ & \\
  40 & $(1.83 \pm 0.50) \times 10^{-5}$ & $1.17 \times 10^{-5}$, RC $\ge 0$ \\
  50 & $(8.54 \pm 2.58) \times 10^{-6}$ & \\
  60 & $(4.59 \pm 1.54) \times 10^{-6}$ & $3.80 \times 10^{-6}$, RC $\ge 1$ \\
  70 & $(2.72 \pm 1.00) \times 10^{-6}$ & \\
  80 & $(1.73 \pm 0.69) \times 10^{-6}$ & \\
  90 & $(1.16 \pm 0.49) \times 10^{-6}$ & \\
 100 & $(8.09 \pm 3.67) \times 10^{-7}$ & $9.16 \times 10^{-7}$, RC $\ge 2$ \\
\enddata
\tablecomments{Abell RC $\ge 0, 1$ densities from \citet{dp02};
RC $\ge 2$ density from a compilation of Abell redshifts by Postman.}
\end{deluxetable}

\end{document}